# A Game Theoretic Approach to Distributed Opportunistic Scheduling

Albert Banchs, Andrés García-Saavedra, Pablo Serrano and Joerg Widmer

***Abstract*—Distributed Opportunistic Scheduling (DOS) is inherently harder than conventional opportunistic scheduling due to the absence of a central entity that has knowledge of all the channel states. With DOS, stations contend for the channel using random access; after a successful contention, they measure the channel conditions and only transmit in case of a good channel, while giving up the transmission opportunity when the channel conditions are poor. The distributed nature of DOS systems makes them vulnerable to *selfish* users: by deviating from the protocol and using more transmission opportunities, a selfish user can gain a greater share of the wireless resources at the expense of the well-behaved users. In this paper, we address the selfishness problem in DOS from a game theoretic standpoint. We propose an algorithm that satisfies the following properties: ($i$) when all stations implement the algorithm, the wireless network is driven to the optimal point of operation, and ($ii$) one or more selfish stations cannot gain any profit by deviating from the algorithm. The key idea of the algorithm is to react to a selfish station by using a more aggressive configuration that (indirectly) punishes this station. We build on *multivariable control theory* to design a mechanism for punishment that on the one hand is sufficiently severe to prevent selfish behavior while on the other hand is light enough to guarantee that, in the absence of selfish behavior, the system is stable and converges to the optimum point of operation. We conduct a game theoretic analysis based on *repeated games* to show the algorithm's effectiveness against selfish stations. These results are confirmed by extensive simulations.**

## I. Introduction

Opportunistic scheduling techniques have been shown to provide substantial performance improvements in wireless networks. These techniques take advantage of the fluctuations in the channel conditions of the different wireless stations; by selecting the station with the best instantaneous channel for data transmission, opportunistic scheduling can utilize wireless resource more efficiently. A key assumption of most opportunistic scheduling techniques [1], [2] is that the scheduler is centralized and has knowledge of the instantaneous channel conditions of all stations.

Distributed Opportunistic Scheduling (DOS) techniques [3]–[6] have been proposed only recently. In contrast to centralized schemes, with DOS each station has to make scheduling decisions without knowledge of the channel conditions of the other stations. Stations contend for the channel using random access with a given access probability. After successful contention, a station measures the channel and, in case of poor channel conditions (i.e., when the instantaneous transmission rate is below a given threshold), the station gives up the transmission opportunity. This allows all stations to recontend for the channel, letting a station with better conditions win the contention, which increases the overall throughput. In this way, DOS techniques exploit both multi-user diversity across stations and time diversity across slots.

The absence of global channel information makes DOS systems very vulnerable to *selfish* users. By deviating from the above protocol and using a more aggressive configuration, a selfish user can easily gain a greater share of the wireless resources at the expense of the other, well-behaved users. In this paper, we address the selfishness problem in DOS from a game theoretic standpoint. In our formulation of the problem, the players are the wireless stations that implement DOS and strive to obtain as much resources as possible from the wireless network. We show that, in the absence of penalties, the wireless network naturally tends to either great unfairness or network collapse. Following this result, we design a penalty mechanism in which any player who misbehaves will be punished by other players in such a way that there is no incentive to misbehave. A key challenge when designing such a penalty scheme is to carefully adjust the punishment inflicted upon a misbehaving station. On the one hand side, if the punishment is too light, a selfish station may still benefit from misbehaving. On the other hand, an overreaction may itself be interpreted as misbehavior and could trigger punishment by other stations, leading to an endless spiral of increasing punishments and a throughput collapse. Addressing this challenge through a combination of game theory and multivariable control theory is a key part of our design.

The most relevant prior work on DOS by Zheng et al. [3] sets the basic foundations of distributed opportunistic scheduling. The authors propose a mechanism based on optimal stopping theory and analyze its performance both with well-behaved and with selfish users. The aim of the algorithm is to maximize the total throughput of the network. [4]–[6] extend the basic mechanism of [3] by analyzing the case of imperfect channel information [4], improving channel estimation through two-level channel probing [5], and incorporating delay constraints [6]. While our algorithm deals with the basic DOS mechanism of [3], it could be extended with the enhancements of [4]–[6]. The key contributions of our work are:

1) We perform a joint optimization of both the transmission rate thresholds and the access probabilities, while [3] only optimizes the thresholds.
2) We provide a proportionally fair allocation that achieves a good tradeoff between total throughput and fairness, while [3] maximizes the total throughput of the network, which may lead to starvation of the stations with poor

A. Banchs is with the University Carlos III of Madrid and with the Institute IMDEA Networks. A. García-Saavedra and P. Serrano are with the University Carlos III of Madrid. J. Widmer is with the Institute IMDEA Networks.

channel conditions.

3) We propose a simple algorithm based on control theory that guarantees stability and quick convergence to the optimal point of operation, in contrast to the comparatively complex heuristics of [3].
4) Our game theoretic analysis considers that users can selfishly configure both their access probability and transmission rate threshold, while the analysis of [3] assumes that selfish users can only maliciously configure the thresholds.
5) We use a penalty mechanism to force an optimal Nash equilibrium, while [3] introduces a pricing mechanism for this purpose, which may not be practical in many scenarios; additionally, the performance of the pricing mechanism heavily depends on the cost parameter and even in the best case is only suboptimal.

The remainder of the paper is organized as follows. In Section II we present an analysis of our system and derive the optimal configuration of access probabilities and transmission rate thresholds. In Section III we show that, in the absence of penalties, the wireless network tends to a highly undesirable resource allocation; based on this, we propose an algorithm named *Distributed Opportunistic scheduling with distributed Control* (DOC) that avoids this situation by implementing a decentralized penalty mechanism that controls selfish behavior through punishments. Section IV shows by means of control theory, that when all the stations implement DOC, the system stably converges to the optimal point of operation obtained in Section II. In Section V we conduct a game theoretic analysis of DOC to show that stations cannot gain any profit from behaving selfishly. The performance of the proposed scheme is extensively evaluated through simulations in Section VI. Finally, Section VII provides some concluding remarks.

## II. Analysis and Optimal configuration

In the following we present our system model and analyze the throughput as a function of the access probabilities and transmission rate thresholds. We then compute the optimal configuration of these parameters for a proportionally fair throughput allocation, which is well known to provide a good tradeoff between total throughput and fairness [7].

### A. System Model

Our system model follows that of [3]–[6]. We consider a single-hop wireless network with $N$ stations, where station $i$ contends for the channel with an access probability $p_i$. A collision model is assumed for the channel access, where the channel contention of a station is successful if no other station contends at the same time. Let $\tau$ denote the duration of a mini slot for channel contention, which can either be empty, or can contain a successful contention or a collision.

As in [3]–[6], we assume that a station $i$ obtains its local channel conditions after a successful contention. Let $R_i(\theta)$ denote the corresponding transmission rate at time $\theta$. If $R_i(\theta)$ is small (indicating a poor channel), station $i$ gives up on this transmission opportunity and lets all the stations recontend. Otherwise, it transmits for a duration of $\mathcal{T}$. Fig. 1 depicts an example of such channel contention. Our model, like that of [3]–[6], assumes that $R_i(\theta)$ remains constant for the duration of a data transmission and that different observations of $R_i(\theta)$ are independent.[1] From [3], we have that the optimal transmission policy is a threshold policy: for a given threshold $\bar{R}_i$, station $i$ only transmits after a successful contention if $R_i(\theta) \geq \bar{R}_i$.

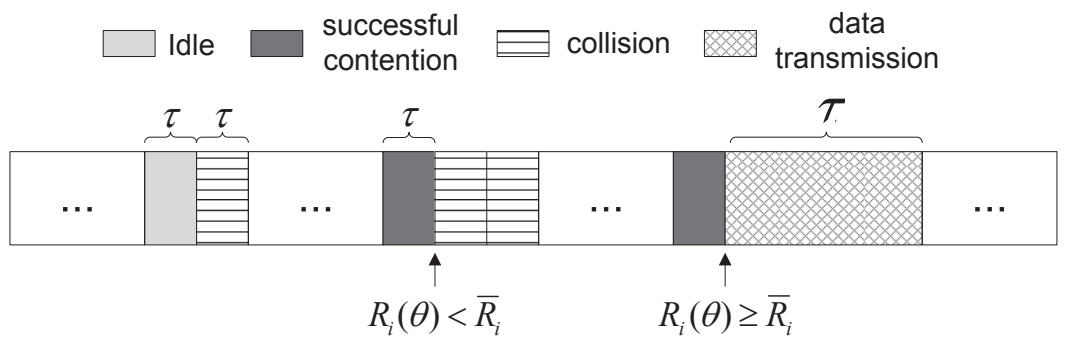

Fig. 1. Channel contention example.

### B. Throughput Analysis

The throughput $r_i$ achieved by station $i$ is a function of the parameters $p_i$ and $\bar{R}_i$. Let $l_i$ be the average number of bits that station $i$ transmits upon a successful contention and $T_i$ be the average time it holds the channel. Then, the throughput of station $i$ is

$$r_i = \frac{p_{s,i} l_i}{\sum_j p_{s,j} T_j + (1 - p_s)\tau} \tag{1}$$

where $p_{s,i}$ is the probability that a mini slot contains a successful contention of station $i$

$$p_{s,i} = p_i \prod_{j \neq i} (1 - p_j) \tag{2}$$

and $p_s$ is the probability that it contains any successful contention

$$p_s = \sum_i p_{s,i} \tag{3}$$

Both $l_i$ and $T_i$ depend on $\bar{R}_i$. Upon a successful contention, a station holds the channel for a time $\mathcal{T} + \tau$ in case it transmits data and $\tau$ in case it gives up the transmission opportunity. Thus, $T_i$ can be computed as

$$T_i = Prob(R_i(\theta) < \bar{R}_i)\tau + Prob(R_i(\theta) \geq \bar{R}_i)(\mathcal{T} + \tau) \tag{4}$$

In case the station uses the transmission opportunity, it transmits a number of bits given by $R_i(\theta)T_i$, which yields

$$l_i = \int_{\bar{R}_i}^{\infty} r T_i f_{R_i}(r) dr \tag{5}$$

where $f_{R_i}(r)$ is the pdf of $R_i(\theta)$.

With the above, we can compute $r_i$ from $\mathbf{p} = \{p_1, \ldots, p_N\}$ and $\bar{\mathbf{R}} = \{\bar{R}_1, \ldots, \bar{R}_N\}$. In the following, we obtain the optimal configuration of these parameters to provide proportional fairness.

[1] The assumption that $R_i(\theta)$ remains constant during a data transmission is a standard assumption for the block-fading channel in wireless communications [8], [9], while the assumption that different observations are independent is justified in [3] through numerical calculations.

### C. Optimal $p_i$ configuration

We start by computing the optimal configuration of $p_i$. Let us define $w_i$ as

$$w_i = \frac{p_{s,i}}{p_{s,1}} \tag{6}$$

where we take station 1 as reference. From the above equation we have that $p_{s,i} = w_i p_s / \sum_j w_j$ and substituting this in Eq. (1) yields

$$r_i = \frac{w_i p_s l_i}{\sum_j w_j p_s T_j + \sum_j w_j (1-p_s)\tau} \tag{7}$$

In a slotted wireless system such as the one of this paper, the optimal success probability is approximately $1/e$ [10]. The problem of finding the $\mathbf{p}$ configuration that maximizes the proportionally fair rate allocation is thus equivalent to finding the values $w_i$ that maximize $\sum_i log(r_i)$ given that $p_s = 1/e$. To obtain these $w_i$ values, we impose

$$\frac{\partial \sum_i log(r_i)}{\partial w_i} = 0 \tag{8}$$

which yields

$$\frac{1}{w_i} - N \frac{p_s T_i + (1-p_s)\tau}{\sum_i w_i p_s T_i + \sum_j w_j (1-p_s)\tau} = 0 \tag{9}$$

Combining this expression for $w_i$ and $w_j$, we obtain

$$\frac{w_i}{w_j} = \frac{p_s T_j + (1-p_s)\tau}{p_s T_i + (1-p_s)\tau} \tag{10}$$

From the above, the solution to the optimization problem is given by the values of $\mathbf{p}$ resulting from solving the following system of equations:

$$\sum_i p_i \prod_{j \neq i} (1-p_j) = \frac{1}{e} \tag{11}$$

$$\frac{p_i \prod_{j\neq i} 1-p_j}{p_1 \prod_{j \neq 1} 1-p_j} = \frac{T_1 + \tau(1/p_s - 1)}{T_i + \tau(1/p_s - 1)}, \; i = 2, \ldots, N \tag{12}$$

This system of equations has two solutions, since $1/e$ is only an approximation to the truly optimal success probability. For one of the solutions, all of the access probabilities are larger than the corresponding ones from the other. We select the solution with the larger access probabilities, denoted by $\mathbf{p}^* = \{p_1^*, \ldots, p_N^*\}$, and refer to them as the *optimal access probabilities*.

Note that determining $\mathbf{p}^*$ above requires computing $T_i \; \forall i$, which depend on the optimal configuration of the thresholds $\bar{\mathbf{R}}$. In the following section we address the computation of the optimal $\bar{\mathbf{R}}$, which we denote by $\bar{\mathbf{R}}^* = \{\bar{R}_1^*, \ldots, \bar{R}_N^*\}$.

### D. Optimal $\bar{R}_i$ configuration

In order to obtain the optimal configuration of $\bar{\mathbf{R}}$, we need to find the transmission threshold of each station that, given the $\mathbf{p}^*$ computed above, optimizes the overall performance in terms of proportional fairness. This is given by the following theorem.

**Theorem 1.** *Let us consider that station $k$ is alone in the channel and it contends for the channel with $p_k = 1/e$. Let $\bar{R}_k^1$ be the transmission rate threshold that optimizes the throughput of this station under the assumption that different channel observations are independent. Then, $\bar{R}_k^* = \bar{R}_k^1$.*

*Proof:* The proof is by contradiction. Assume that there exists a configuration $\bar{\mathbf{R}}^*$ with $\bar{R}_k^* \neq \bar{R}_k^1$ for some station $k$ that provides proportional fairness.

Let $l_k^1$ and $T_k^1$ be the values of $l_k$ and $T_k$ for the threshold $\bar{R}_k^1$ and $l_k^*$ and $T_k^*$ the corresponding values for $\bar{R}_k^*$. Since $\bar{R}_k^1$ maximizes $r_k$ when station $k$ is alone:

$$\frac{l_k^1}{T_k^1 + (e-1)\tau} > \frac{l_k^*}{T_k^* + (e-1)\tau} \tag{13}$$

Let us consider that there are $N$ stations in the network and the configuration $\bar{\mathbf{R}}^*$ is used. Given $\bar{\mathbf{R}}^*$, the $\mathbf{p}^*$ that maximizes $\sum_i log(r_i)$ is given by Eqs. (11) and (12). This leads to the following throughput for station $k$:

$$r_k^* = \frac{p_{s,k}^* l_k^*}{\sum_j p_{s,j}^* (T_j^* + (e-1)\tau)} = \frac{l_k^*}{N(T_k^* + (e-1)\tau)} \tag{14}$$

and for the other stations:

$$r_i^* = \frac{l_i^*}{N(T_i^* + (e-1)\tau)} \;, \; \forall i \neq k \tag{15}$$

Let us now consider the alternative configuration $\bar{R}_k^1$ for station $k$ and $\bar{R}_i^*$ for the other stations. Let us take the $p_k^1$ and $p_i^1$ configuration that satisfies Eqs. (11) and (12) with this alternative configuration. This yields the following throughput for station $k$:

$$r_k^1 = \frac{l_k^1}{N(T_k^1 + (e-1)\tau)} > r_k^* \tag{16}$$

and for the other stations:

$$r_i^* = \frac{l_i^*}{N(T_i^* + (e-1)\tau)} \;, \; \forall i \neq k \tag{17}$$

With the above, we have found an alternative configuration that provides a higher throughput to station $k$ and the same throughput to all other stations. Therefore, this alternative configuration increases $\sum_i log(r_i)$, which contradicts the initial assumption that the configuration $\bar{\mathbf{R}}^*$ provides proportional fairness. ∎

Following the above theorem, the optimal configuration of the thresholds $\bar{\mathbf{R}}^*$ can be computed based on *optimal stopping theory*. This is done in [3] which finds that the optimal threshold $\bar{R}_i^*$ can be obtained by solving the following fixed point equation:

$$E\left(R_i(\theta) - \bar{R}_i^*\right)^+ = \frac{\bar{R}_i^* \tau}{\mathcal{T}/e} \tag{18}$$

The above concludes the search for the optimal configuration. The key advantage of this configuration is that it allows each station to compute its $\bar{R}_i^*$ based on *local information* only, which decouples the computation of $\bar{R}_i^*$ from that of $p_i^*$. Based on this finding, we now present a distributed mechanism to compute the optimal configuration where each station uses a fixed $\bar{R}_i = \bar{R}_i^*$ obtained locally, together with an adaptive algorithm to determine the optimal $p_i^*$.

## III. DOC ALGORITHM

In this section we propose an adaptive algorithm that satisfies the following properties: $(i)$ when all stations implement the algorithm, it leads to the optimal configuration computed above, and $(ii)$ a selfish station cannot gain profit by deviating from the algorithm. We first motivate our algorithm by showing that, in the absence of punishments, the system will naturally tend to a highly undesirable point of operation. Then, we present our algorithm which uses punishments to drive the system to the optimal point of operation obtained in the previous section.

### A. Motivation

If no constrains are imposed on the wireless network and stations are allowed to configure their $\{p_i, \bar{R}_i\}$ parameters to selfishly maximize their profit, the network will not naturally tend to the optimum configuration above. In order to show this, we model the wireless system as a static game in which each station can choose its configuration without suffering any penalty. The following theorem characterizes the Nash equilibria of this game.

**Theorem 2.** *In absence of penalties, there is at least one station that plays $p_i = 1$ in any Nash equilibrium.*

*Proof:* The proof is by contradiction. Assume that there is a Nash equilibrium such that $p_j \neq 1\ \forall j$. Now take one player $i$ with throughput

$$r_i = \frac{p_i \prod_{j \neq i} (1 - p_j) l_i}{p_i \hat{T}_i + (1 - p_i)\hat{T}_{-i}} \tag{19}$$

where $\hat{T}_i$ is the average duration the channel is occupied when station $i$ transmits and $\hat{T}_{-i}$ is the average duration of a transmission or an empty mini slot when station $i$ does not transmit.

Taking the partial derivative we have

$$\frac{\partial r_i}{\partial p_i} = \frac{\prod_{j \neq i} (1 - p_j) l_i \hat{T}_{-i}}{\left(p_i \hat{T}_i + (1 - p_i)\hat{T}_{-i}\right)^2} > 0 \tag{20}$$

It can be seen that the throughput $r_i$ is a strictly increasing function of $p_i$ (given that the $\bar{R}_i$ configuration as well as the configuration of the other stations does not change).

From the above follows that $\{p_i, \bar{R}_i\}$, with $p_i \neq 1$, is not the best strategy for player $i$ given the configuration of the other stations, since $i$ would obtain a higher throughput for $p_i = 1$ and the same $\bar{R}_i$. Thus, this solution is not a Nash equilibrium, which contradicts our initial assumption. ■

Any of the above Nash equilibria are highly undesirable. If station $i$ is the only one that plays $p_i = 1$, then player $i$ achieves non-zero throughput while all other players have zero throughput. Conversely, any other station $j$ also playing $p_j = 1$ results in a network collapse and all players obtain zero throughput.

We conclude from the above that, in the absence of punishments, selfish behaviors will severely degrade the performance of the wireless system. In the following, we propose an algorithm that addresses this problem by implementing a distributed punishment mechanism.

### B. Rationale behind the algorithm

Before presenting the algorithm, we first discuss the rationale that lies behind its design. This rationale heavily relies on the notion of *channel time* that a station obtains over a certain interval, defined as

$$t_i = \sum_{j=1}^{n_i} (T_i(j) + (e - 1)\tau) \tag{21}$$

where $n_i$ is the number of successful contentions of station $i$ in that period and $T_i(j)$ is the duration of the $j^{th}$ successful contention of the station. The above definition comprises the aggregated transmission time of the station plus a fixed overhead of $(e - 1)\tau$ that is added every time the station accesses the channel.

An important observation that drives the design of our algorithm is that, with the configuration of Section II, all stations receive the same channel time, i.e., $t_i = t_j\ \forall i, j$. This can be seen as follows. From Eq. (21) we have that over a given interval,

$$\frac{t_i}{t_j} = \frac{n_i\,(T_i + (e - 1)\tau)}{n_j\,(T_j + (e - 1)\tau)} = \frac{p_{s,i}(T_i + (e - 1)\tau)}{p_{s,j}(T_j + (e - 1)\tau)} \tag{22}$$

since by definition $n_i/n_j = p_{s,i}/p_{s,j}$. Furthermore, from Eq. (12) we have $p_{s,i}(T_i + (e - 1)\tau) = p_{s,j}(T_j + (e - 1)\tau)$ and thus $t_i = t_j$.

Since the overhead in the definition of channel time, $(e - 1)\tau$, coincides with the average time between two successes for the optimal configuration, from the above follows that, when all stations use the optimal configuration over a given time interval $T_{total}$, they all observe the same *optimal channel time* $t^*$,

$$t^* = T_{total}/N \tag{23}$$

The last observation upon which our algorithm relies is that as long as a selfish station does not receive more channel time than $t^*$, it cannot increase its throughput. The throughput of a station with a given channel time and $\bar{R}_i$ is equal to the throughput it would obtain if it were alone in the channel during this time with $p_i = 1/e$ and the same $\bar{R}_i$. From Theorem 1, we have that this throughput is maximized for the optimal transmission rate threshold $\bar{R}_i^*$. Therefore, as long as the station does not receive extra channel time, it will not be able to achieve a higher throughput.

Given these observations, we base our algorithm on the following principles: $(i)$ if a given station $i$ detects that another station $k$ is receiving a larger channel time than itself, then station $k$ is considered selfish and punished by station $i$, and $(ii)$ when punishing station $k$, station $i$ needs to make sure the punishment is severe enough so that station $k$'s channel time remains below $t^*$ and thus it cannot benefit from misbehaving.

### C. Algorithm design

The DOC algorithm aims at driving the system to the optimal configuration $\{\mathbf{p}^*, \bar{\mathbf{R}}^*\}$ obtained in Section II. As discussed in Section II-D, the optimal configuration of $\bar{R}_i$ can be computed locally by each station independently of the other stations. Therefore, with DOC each station maintains a fixed $\bar{R}_i$ (equal to the optimal value) and implements an adaptive algorithm to configure its access probability $p_i$.

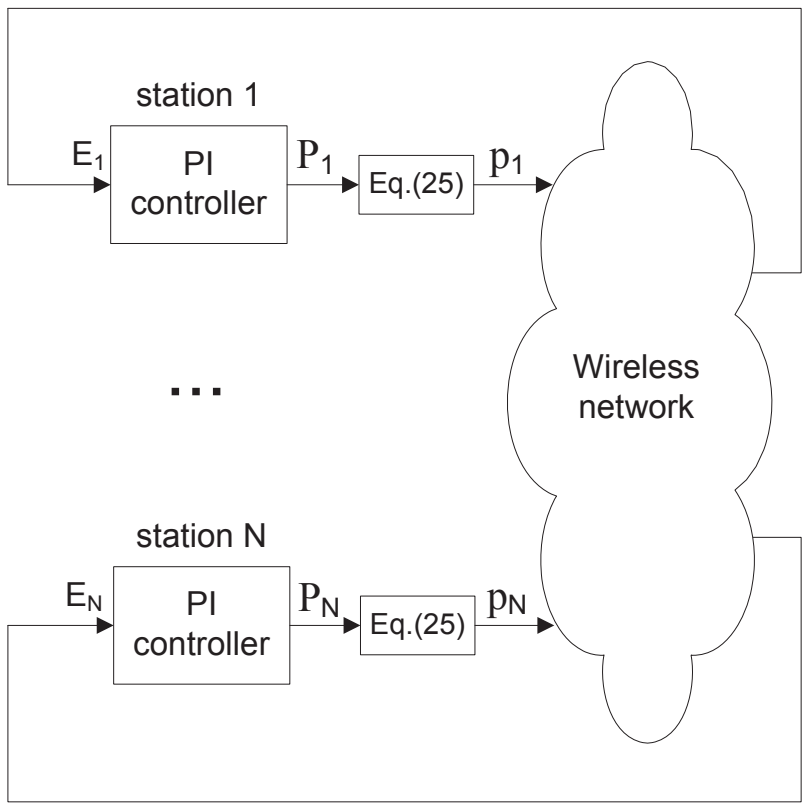

Fig. 2. DOC control system.

Time is divided into intervals in such a way that each station updates its access probability $p_i$ at the beginning of an interval. The central idea behind DOC is that when a misbehaving station is detected, the other stations increase their access probabilities in subsequent intervals to prevent the selfish station from benefiting from the misbehavior.

A key challenge in DOC is to carefully adjust the reaction against a selfish station. If the reaction is not severe enough, a selfish station may benefit from misbehaving, but if the reaction is too severe, the system may turn unstable by entering an endless loop where all stations indefinitely increase their $p_i$ to punish each other.

Control theory is a particularly suitable tool to address this challenge, since it helps to guarantee the convergence and stability of adaptive algorithms. We use techniques from *multi-variable control theory* [11] for the design of the DOC algorithm. The algorithm is based on the classic system illustrated in Fig. 2, where each station runs an independent controller in order to compute its configuration. The controller that we have chosen for this paper is a *proportional-integral* (PI) controller, a well known controller from classic control theory that has been used by a number of networking algorithms in the literature [12]–[14].

As shown in the figure, the PI controller of station $i$ takes the error signal $E_i$ as input and provides the control signal $P_i$ as output. The error signal serves to evaluate the state of the system. If the system is operating as desired, the error signal of all stations is zero. Otherwise, the error is non-zero and we need to drive the system from its current state to the desired point of operation. In order to do this, the PI controller adjusts the control signal $P_i$ by (appropriately) increasing it when $E_i > 0$ and decreasing it otherwise. In the following, we address the design of $P_i$ and $E_i$.

### D. Control signal $P_i$

The goal of the adaptive algorithm implemented by the controller of a station is to adjust the access probability $p_i$ with which the station contends. Hence, there needs to be a one-to-one mapping between the control signal $P_i$ and $p_i$. In addition, we impose that in the optimal point of operation, the $P_i$ values of all stations are the same. This latter requirement is necessary to obtain the conditions for stability in Section IV.

Based on the above requirements, we design $P_i$ as

$$P_i = \frac{p_i}{1-p_i}\left(T_i + (e-1)\tau\right) \tag{24}$$

Hence, a station can compute its $p_i$ from the control signal $P_i$ as

$$p_i = \frac{P_i}{T_i + (e-1)\tau + P_i} \tag{25}$$

### E. Error signal $E_i$

The design of the error signal $E_i$ has the following two goals: ($i$) selfish stations should not be able to obtain extra channel time from the wireless network by using a configuration different from the optimal one, and ($ii$) as long as there are no selfish stations, $\mathbf{p}$ should converge to the optimal $\mathbf{p}^*$.

To this end, each station measures its channel time as well as that of the other stations at the end of every interval and computes the error signal

$$E_i = \sum_{j \neq i} (t_j - t_i) - F_i \tag{26}$$

where $F_i$ is a function that we design below. The error signal $E_i$ consists of the two components:

- The first component ($\sum_{j \neq i} t_j - t_i$) punishes selfish stations. If a station $i$ receives less channel time than the other stations, this component will be positive and hence station $i$ will increase its access probability $p_i$.
- The second component ($F_i$) drives the system to the desired point of operation in the absence of selfish behavior (i.e., when all stations receive the same channel time).

We next address the design of the function $F_i$. In order to drive the current $\mathbf{p}$ to the desired $\mathbf{p}^*$ when $t_i = t_j \; \forall i, j$, we need $F_i > 0$ for $p_i > p_i^*$, such that in this case $p_i$ decreases, and $F_i < 0$ for $p_i < p_i^*$.

Another requirement when designing $F_i$ is that selfish stations should not be able to obtain more channel time than $t^*$. We first consider the case where all stations are well-behaved and run the DOC algorithm except one that is selfish. In this case, the error signal allows that the selfish station obtains an additional channel time equal to $F_i$: taking $E_i = 0$ and setting $t_i = t$ for all stations but the selfish one, the channel time $t_k$ of the selfish station is given as

$$t_k = t + F_i \tag{27}$$

As argued before, $F_i$ needs to be small enough such that

$$t_k \leq t^* \tag{28}$$

Combining the two equations above yields

$$t_k + (N-1)t + (N-1)F_i \leq Nt^* \tag{29}$$

Using $\sum_j t_j = t_k + (N-1)t$ we can isolate $F_i$ and obtain

$$F_i \leq \frac{1}{N-1} D \tag{30}$$

where $D$ is defined as[2]

$$D = Nt^* - \sum_j t_j \tag{31}$$

With this constraint on $F_i$, the additional channel time obtained by a selfish station does not compensate for the overall efficiency loss due to sub-optimal access probabilities, and hence a selfish station cannot benefit from misbehaving.

In addition, multiple selfish stations should not be able to gain any aggregated channel time by playing a coordinated strategy. We consider $m$ selfish stations and again set the $t_i$ of the other stations equal to $t$. From $E_i = 0$ we have

$$\sum_{j=1}^{m} t_j = mt + F_i \tag{32}$$

and we require that

$$\sum_{j=1}^{m} t_j \leq mt^* \tag{33}$$

Combining the above equations and isolating $F_i$ yields

$$F_i \leq \frac{m}{N-m} D \tag{34}$$

Eqs. (30) and (34) provide the maximum value for $F_i$ that still prevents one or more selfish stations to benefit from misbehaving. Given all these requirements, we design $F_i$ as:

$$F_i = \begin{cases} \min((N-1)D, D/N), & p_i > p_i^{min} \\ \min((N-1)D, -D/N, (N-1)\Delta), & p_i \leq p_i^{min} \end{cases} \tag{35}$$

where $\mathbf{p^{min}} = \{p_1^{min}, \ldots, p_N^{min}\}$ are the access probabilities that minimize $D$ subject to $t_i = t_j \; \forall i, j$ and $\Delta$ is the value that $D$ takes at this point,

$$\Delta = D|_{\mathbf{p}=\mathbf{p^{min}}} \tag{36}$$

The above design satisfies Eqs. (30) and (34) and fulfills $F_i > 0$ for $p_i > p_i^*$ and $F_i < 0$ for $p_i < p_i^*$ (when $t_i = t_j \; \forall i, j$). It thus meets all the requirements set above for function $F_i$. Note that the term $D/N$ ensures that Eqs. (30) and (34) are satisfied when $D > 0$, the term $(N-1)D$ ensures that they are satisfied when $D < 0$, and the terms $(N-1)\Delta$ and $-D/N$ ensure that $F_i < 0$ when $t_i = t_j \; \forall i, j$ and $p_i < p_i^*$, as illustrated in Fig. 3.

Note that we keep $F_i$ very close to the upper bound for $p_i > p_i^*$, which means that the degree of punishment inflicted upon selfish stations is as small as the above requirements allow. The rationale for this design is that any punishment in the form of an increase in access probabilities affects selfish stations and well-behaved stations alike. Providing just enough punishment to prevent any throughput gain for selfish stations maintains the highest level of overall throughput for the system in the presence of malicious users or in transient conditions. Following a similar rationale, we keep $F_i$ well below the upper bound for $p_i < p_i^*$.

[2]Note that $D$ is a function of the efficiency in channel contention, which depends on $\mathbf{p}$: if channel contention is more efficient, we have a larger number of data transmissions in the interval, which results in a larger sum of channel times and therefore a smaller $D$.

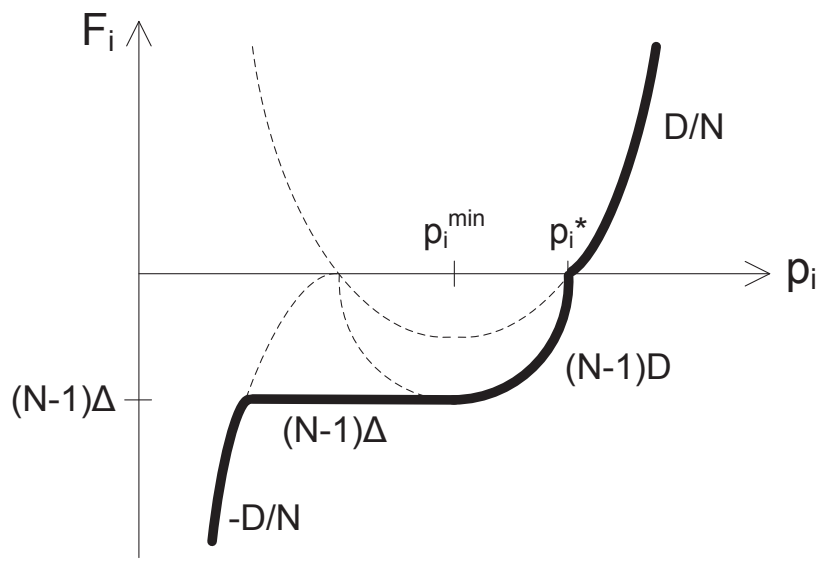

Fig. 3. $F_i$ as a function of $p_i$ when $t_i = t_j \; \forall i, j$.

This concludes the design of the algorithm. In the following two sections, we analytically evaluate its performance when all stations are well-behaved (Section IV) and when some of them behave selfishly (Section V).

## IV. DOC Analysis

We first analyze the wireless system under steady state conditions and show that it is driven to the desired point of operation obtained in Section II. We then conduct a transient analysis and derive the sufficient conditions for stability.

### A. Steady state analysis

Since the controller includes an integrator, there is no steady state error [15] and the steady solution can be obtained from

$$E_i = 0 \; \forall i \tag{37}$$

Using Eqs. (26) and (35), $E_i$ can be computed from $t_i$ and $t^*$, which allows expressing Eq. (37) as a system of equations of $\mathbf{p}$. The following theorem guarantees the uniqueness of this system of equations and shows that the unique stable point in steady state is the desired point of operation from Section II.[3]

**Theorem 3.** *The unique stable point of operation of the system in steady state is* $\mathbf{p} = \mathbf{p}^*$.

*Proof:* Let us consider two stations $i$ and $j$. From Eq. (37) we have $E_i - E_j = 0$, which yields

$$Nt_j + F_j - Nt_i - F_i = 0 \tag{38}$$

Note that $t_j > t_i$ implies $F_j \geq F_i$, and vice versa. Therefore, the above requires that $t_i = t_j \; \forall i, j$. Substituting this into $E_i = 0$ yields $F_i = 0$. Given $t_i = t_j$, $F_i$ is an increasing function of $p_i$ that crosses 0 at $p_i = p_i^*$. Hence, the only $p_i$ that satisfies $F_i = 0$ is $p_i^*$. Since this holds for all $i$, the unique stable point of operation is $p_i = p_i^* \; \forall i$. ∎

### B. Stability analysis

We next conduct a stability analysis of DOC to configure the parameters of the PI controller. Following the definition of a PI controller [15], station $i$ computes the value of $P_i$ at interval $\Theta'$ as a function of the error values measured by

[3]While the existence of a unique point of operation can be easily guaranteed in a centralized system where the configuration of all stations is imposed by a central entity, it is much harder to guarantee in a distributed system in which each station chooses its own configuration.

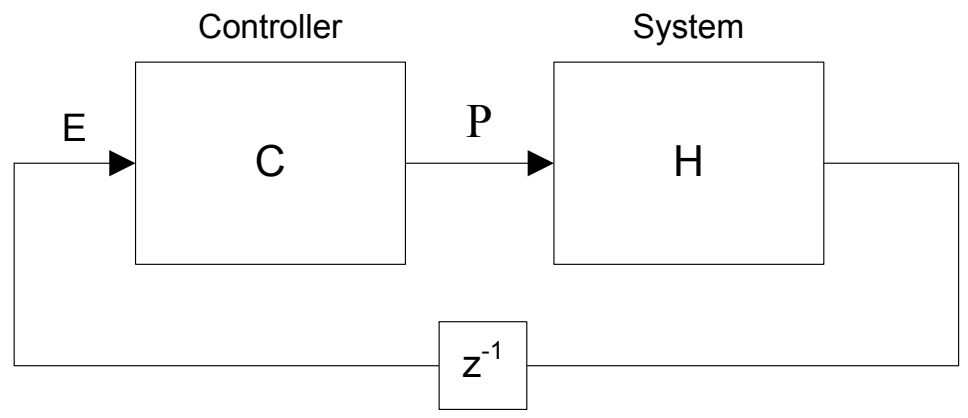

Fig. 4. Control system

the station in the current and previous intervals based on the following equation:

$$P_i(\Theta') = K_p E_i(\Theta') + K_i \sum_{\Theta=0}^{\Theta'-1} E_i(\Theta) \tag{39}$$

where $K_p$ and $K_i$ are the parameters of the controller that we have to configure.

The DOC system shown in Fig. 2 can be expressed in the form of Fig. 4. In this figure, $C$ represents the function implemented by the controllers, which computes the control signals $P_i$ taking as input the error signals $E_i$, and $H$ represents the wireless system which provides the error signals $E_i$ measured by the stations based on the control signals $P_i$. The control and error signals in the figure are given by the following vectors:

$$\mathbf{P} = (P_1, \ldots, P_N)^T \tag{40}$$

and

$$\mathbf{E} = (E_1, \ldots, E_N)^T \tag{41}$$

Our control system consists of one PI controller in each station $i$ that takes $E_i$ as input and gives $P_i$ as output. Following this, we can express the relationship between $\mathbf{E}$ and $\mathbf{P}$ as follows

$$\mathbf{P}(z) = C \cdot \mathbf{E}(z) \tag{42}$$

where

$$C = \begin{pmatrix} C_{PI}(z) & 0 & 0 & \ldots & 0 \\ 0 & C_{PI}(z) & 0 & \ldots & 0 \\ 0 & 0 & C_{PI}(z) & \ldots & 0 \\ \vdots & \vdots & \vdots & \ddots & \vdots \\ 0 & 0 & 0 & \ldots & C_{PI}(z) \end{pmatrix} \tag{43}$$

with $C_{PI}(z)$ being the z transform of a PI controller [15],

$$C_{PI}(z) = K_p + \frac{K_i}{z-1} \tag{44}$$

In order to analyze our system from a control theoretic standpoint, we need to characterize the wireless system with a transfer function $H$ that takes $\mathbf{P}$ as input and has $\mathbf{E}$ as output. Eq. (26) gives a nonlinear relationship between $\mathbf{E}$ and $\mathbf{P}$. In order to express this relationship as a transfer function, we linearize it when the system suffers small perturbations around its stable point of operation. We then study the linearized model and force that it is stable. Note that the stability of the linearized model guarantees that our system is locally stable.[4]

[4]A similar approach was used in [16] to analyze RED from a control theoretic standpoint.

We express the perturbations around the stable point of operation as follows:

$$\mathbf{P} = \mathbf{P}^* + \delta\mathbf{P} \tag{45}$$

where $\mathbf{P}^*$ is the stable point of operation as given by Eq. (24) with $\mathbf{p} = \mathbf{p}^*$.

With the above, the perturbations suffered by $\mathbf{E}$ can be approximated by

$$\delta\mathbf{E} = H \cdot \delta\mathbf{P} \tag{46}$$

where

$$H = \begin{pmatrix} \frac{\partial E_1}{\partial P_1} & \frac{\partial E_1}{\partial P_2} & \cdots & \frac{\partial E_1}{\partial P_N} \\ \frac{\partial E_2}{\partial P_1} & \frac{\partial E_2}{\partial P_2} & \cdots & \frac{\partial E_2}{\partial P_N} \\ \vdots & \vdots & \ddots & \vdots \\ \frac{\partial E_N}{\partial P_1} & \frac{\partial E_N}{\partial P_2} & \cdots & \frac{\partial E_N}{\partial P_N} \end{pmatrix} \tag{47}$$

In order to compute these partial derivatives we proceed as follows. The error signal $E_i$ can be expressed as

$$E_i = T_{total} \frac{\sum_{j \neq i} (p_{s,j}(T_j + (e-1)\tau) - p_{s,i}(T_i + (e-1)\tau))}{\sum_j p_{s,j} T_j + (1-p_s)\tau} - F_i \tag{48}$$

The above can be rewritten as a function of $\mathbf{P}$ given by

$$E_i = T_{total} \frac{\sum_{j \neq i} (P_j - P_i)}{\sum_j P_j - \frac{p_s}{p_e}(e-1)\tau + \frac{1-p_s}{p_e}\tau} - F_i \tag{49}$$

where $p_e = \prod_i 1 - p_i$.

We start by showing that $\partial F_i / \partial P_i = 0$ at the stable point of operation. It follows from Eq. (35) that

$$\frac{\partial F_i}{\partial P_i} = 0 \iff \frac{\partial D}{\partial P_i} = 0 \tag{50}$$

$D$ can be expressed as

$$D = Nt^* - T_{total} \frac{\sum_i p_{s,i} T_i + p_s(e-1)\tau}{\sum_i p_{s,i} T_i + (1-p_s)\tau} \tag{51}$$

The partial derivative of $D$ can be computed as

$$\frac{\partial D}{\partial P_i} = \frac{\partial D}{\partial p_i} \frac{\partial p_i}{\partial P_i} \tag{52}$$

Taking the partial derivative of Eq. (51) with respect to $p_i$ and evaluating it at the stable point of operation yields

$$\frac{\partial D}{\partial p_i} = T_{total} \left( \frac{e\tau}{\sum_i p_{s,i} T_i + (e-1)\tau} \right) \left( \frac{\partial p_s}{\partial p_i} \right) \tag{53}$$

Since $p_s$ takes a maximum at the stable point of operation, we have that $\partial p_s / \partial p_i = 0$, which yields $\partial D / \partial P_i = 0$ and hence

$$\frac{\partial F_i}{\partial P_i} = 0 \tag{54}$$

The partial derivative of $E_i$ evaluated at the stable point of operation can then be computed from Eq. (49) as

$$\frac{\partial E_i}{\partial P_i} = -(N-1) T_{total} \frac{1}{\sum_j P_j} \tag{55}$$

Following a similar reasoning, it can be seen that

$$\frac{\partial E_i}{\partial P_j} = T_{total} \frac{1}{\sum_j P_j} \tag{56}$$

Substituting these expressions in matrix $H$ gives

$$H = K_H \begin{pmatrix} -(N-1) & 1 & \dots & 1 \\ 1 & -(N-1) & \dots & 1 \\ \vdots & \vdots & \ddots & \vdots \\ 1 & 1 & \dots & -(N-1) \end{pmatrix} \quad (57)$$

where

$$K_H = T_{total} \frac{1}{\sum_j P_j} \quad (58)$$

With the above, we have the linearized system fully characterized by matrices $C$ and $H$. The next step is to configure the $K_p$ and $K_i$ parameters of this system. The following theorem provides the sufficient conditions of $\{K_p, K_i\}$ for stability:

**Theorem 4.** *The linearized system is guaranteed to be stable as long as $K_p$ and $K_i$ meet the following conditions:*

$$K_i < K_p + \frac{1}{NK_H} \quad (59)$$

$$K_i > 2K_p - \frac{1}{NK_H} \quad (60)$$

*Proof:* According to (6.22) of [11], we need to verify that the following transfer function is stable

$$(I - z^{-1}CH)^{-1}C \quad (61)$$

Computing the above matrix yields

$$(I - z^{-1}CH)^{-1}C = \begin{pmatrix} a & b & b & \dots & b \\ b & a & b & \dots & b \\ b & b & a & \dots & b \\ \vdots & \vdots & \vdots & \ddots & \vdots \\ b & b & b & \dots & a \end{pmatrix} \quad (62)$$

where

$$a = \frac{C_{PI}(z)}{N} \left( 1 + \frac{N-1}{1 + Nz^{-1}K_H C_{PI}(z)} \right) \quad (63)$$

$$b = \frac{C_{PI}(z)}{N} \left( 1 - \frac{1}{1 + Nz^{-1}K_H C_{PI}(z)} \right) \quad (64)$$

Rearranging the terms of the above two equations, we obtain

$$a = \frac{P_1(z)}{z^2 + a_1 z + a_2} \quad (65)$$

$$b = \frac{P_2(z)}{z^2 + a_1 z + a_2} \quad (66)$$

where $P_1(z)$ and $P_2(z)$ are polynomials and

$$a_1 = NK_H K_p - 1 \quad (67)$$

$$a_2 = NK_H(K_i - K_p) \quad (68)$$

According to Theorem 3.5 of [11], a sufficient condition for the stability of a transfer function is that the zeros of its pole polynomial (which is the least common denominator of all the minors of the transfer function matrix) fall within the unit circle. Applying this theorem to $(I - z^{-1}CH)^{-1}C$ yields that the roots of the polynomial $z^2 + a_1 z + a_2$ have to fall inside the unit circle. This can be ensured by choosing coefficients $\{a_1, a_2\}$ that satisfy the following three conditions [17]: $a_2 < 1$, $a_1 < a_2 + 1$ and $a_1 > -1 - a_2$. The third condition is satisfied as long as $K_i > 0$, while the other two yield $K_i < K_p + 1/(NK_H)$ and $K_i > 2K_p - 1/(NK_H)$, respectively. ∎

In addition to guaranteeing stability, our goal in the configuration of the $\{K_p, K_i\}$ parameters is to find the right tradeoff between speed of reaction to changes and oscillations under steady conditions. To this end, we use the *Ziegler-Nichols* rules [18], which have been designed for this purpose. First, we compute the parameter $K_u$, defined as the $K_p$ value that leads to instability when $K_i = 0$, and the parameter $T_i$, defined as the oscillation period under these conditions. Then, $K_p$ and $K_i$ are configured as follows:

$$K_p = 0.4 K_u \quad (69)$$

and

$$K_i = \frac{K_p}{0.85 T_i} \quad (70)$$

In order to compute $K_u$ we proceed as follows. From Eq. (59) with $K_i = 0$ we have the following condition for stability

$$K_p < \frac{1}{2NK_H} \quad (71)$$

We take $K_u$ as the value that may turn the system unstable

$$K_u = \frac{1}{2NK_H} \quad (72)$$

and set $K_p$ according to Eq. (69),

$$K_p = \frac{0.4}{2NK_H} \quad (73)$$

With the $K_p$ value that leads to instability, a given set of input values may change their sign at most every interval, yielding an oscillation period of two intervals ($T_i = 2$). Thus, from Eq. (70),

$$K_i = \left( \frac{1}{0.85 \cdot 2} \right) \frac{0.4}{2NK_H} \quad (74)$$

which completes the configuration of the PI controller parameters. The stability of this configuration is guaranteed by the following corollary:

**Corollary 1.** *The $K_p$ and $K_i$ configuration given by Eqs. (73) and (74) is stable.*

*Proof:* It is easy to see that Eqs. (73) and (74) meet the conditions of Theorem 4. ∎

## V. Game Theoretic Analysis

In the previous section we have seen that, when all stations follow the DOC algorithm, they all play with $p_i = p_i^*$ and $\bar{R}_i = \bar{R}_i^*$. In this section we conduct a game theoretic analysis to show that one or more stations cannot gain any profit by deviating from DOC. In what follows, we say that a station is *honest* or well-behaved when it implements the DOC algorithm to configure its $p_i$ and $\bar{R}_i$ parameters, while we say that it is *selfish* or misbehaving when it plays a different strategy from DOC to configure these parameters with the aim of obtaining some gains.

The game theoretic analysis conducted in this section assumes that users are *rational* and want to maximize their own benefit or *utility*, which is given by the throughput. The model is based on the theory of *repeated games* [19]. With repeated games, time is divided into stages and a player can take new decisions at each stage based on the observed behavior of the other players in the previous stages. This matches our algorithm, where time is divided into intervals and stations update their configuration at each interval.[5] Like other previous analyses on repeated games [20], [21], we consider an infinitely repeated game, which is a common assumption when the players do not know when the game will finish.

### A. Single selfish station

While the design of the DOC algorithm in Section III guarantees that a station cannot gain any profit by playing with a *fixed* selfish configuration, selfish stations might still gain by *varying* their configuration over time. As an example, let us consider a naive algorithm that only takes into account the stations' behavior in the previous stage. While this algorithm may be effective against a fixed selfish configuration, it could easily be defeated by a selfish station that alternates a selfish configuration ($p_k = 1, \bar{R}_k = 0$) with an honest one ($p_k = p_k^*, \bar{R}_k = \bar{R}_k^*$) at every other stage. Since this station would play selfish when all the others play honest, it would achieve a significantly higher throughput every other interval, thus benefiting from its misbehavior.

The above example shows that it is important to make sure that a selfish station cannot gain any profit no matter how it varies its configuration over time. The following theorem confirms the effectiveness of DOC against any (fixed or variable) selfish strategy. The proof of the theorem relies on the integrator component of the PI controller, which keeps track of the aggregated channel time received by all stations and can thus be used to guarantee that this aggregate does not exceed a given amount.

**Theorem 5.** *Let us consider a selfish station that uses a $p_k(\Theta)$ and $\bar{R}_k(\Theta)$ configuration that can vary over time. If all the other stations implement the DOC algorithm, the throughput received by this station will be no larger than $r_k^*$ (where $r_k^*$ is the throughput that station $k$ receives when all stations play DOC).*

*Proof:* The PI controller computes $P_i$ at a given interval $\Theta'$ according to the following expression:

$$\begin{aligned} P_i(\Theta') &= P_i^{initial} + K_p \left( \sum_{j \neq i} (t_j(\Theta') - t_i(\Theta')) - F_i(\Theta') \right) \\ &\quad + K_i \sum_{\Theta=0}^{\Theta'} \left( \sum_{j \neq i} (t_j(\Theta) - t_i(\Theta)) - F_i(\Theta) \right) \end{aligned} \quad (75)$$

[5]Note that the game theoretic study conducted in Section III-A was based on static games instead of repeated ones. The reason is that in Section III-A we considered a system without penalties and hence we could model it as a static game where all players only make a single move at the beginning of the game and (as they are never penalized) do not need to make any further move during the rest of the game.

Note that with the above, $P_i$ will stay between 0 and a given maximum value $P_i^{max}$. If at some time $P_i$ reaches a $P_i^{max}$ value such that $p_i = 1$, then we have $t_j = 0$ for $j \neq i$ and $F_i > -(N-1)t_i$, which yields $E_i < 0$ and therefore $P_i$ decreases. Similarly, if at some time $P_i$ reaches 0, then $t_i = 0$ and $F \leq 0$, which yields $E_i > 0$ and therefore $P_i$ increases.

Considering that $0 \leq P_i(\Theta') \leq P_i^{max}$, the above equation can be expressed as

$$\sum_{\Theta} \left( \sum_{j \neq i} (t_j(\Theta) - t_i(\Theta)) - F_i(\Theta) \right) = K \quad (76)$$

where $K$ is a bounded value.

Let us consider the case in which there is a selfish station that changes its configuration over time and receives a channel time $t_k(\Theta)$ while the other stations are well-behaved and use the same configuration obtaining the same channel time $t(\Theta)$. Then the above can be expressed as

$$\sum_{\Theta} t_k(\Theta) = \sum_{\Theta} (t(\Theta) + F_i(\Theta)) + K \quad (77)$$

Let us consider now a given interval $\Theta$. From Eq. (30) we have

$$F_i(\Theta) \leq \frac{1}{N-1} (t^* - t_k(\Theta) - (N-1)t(\Theta)) \quad (78)$$

which yields

$$(N-1)t(\Theta) + t_k(\Theta) + (N-1)F_i(\Theta) \leq Nt^* \quad (79)$$

Since the above equation is satisfied for all $\Theta$,

$$\sum_{\Theta} (N-1)t(\Theta) + t_k(\Theta) + (N-1)F_i(\Theta) \leq \sum_{\Theta} Nt^* \quad (80)$$

Furthermore, from Eq. (77),

$$(N-1) \sum_{\Theta} t_k(\Theta) = (N-1) \sum_{\Theta} (t(\Theta) + F_i(\Theta)) + (N-1)K \quad (81)$$

Adding the above two equations yields

$$N \sum_{\Theta} t_k(\Theta) \leq N \sum_{\Theta} t^* + (N-1)K \quad (82)$$

from which

$$\sum_{\Theta} t_k(\Theta) \leq \sum_{\Theta} t^* + \frac{N-1}{N} K \quad (83)$$

If we consider a very long period of time, the constant term in the above equation can be neglected and we obtain

$$\sum_{t} t_k(\Theta) \leq \sum_{t} t^* \quad (84)$$

From the above, we have that the selfish station cannot receive more channel time with a selfish strategy than by playing DOC and, following the reasoning of Section III-B, therefore cannot obtain more throughput than it would obtain by playing DOC, i.e.

$$r_k \leq r_k^* \quad (85)$$

which proves the theorem. ■

From the above theorem follows Corollary 2.

**Corollary 2.** *A state in which all stations play DOC (*All-DOC*) is a Nash equilibrium of the game.*

*Proof:* According to Theorem 5, if all stations but one play DOC, then the best response of this station is to play DOC as well since it cannot benefit from playing a different strategy. Thus, *All-DOC* is a Nash equilibrium. ■

This shows that, if all stations start playing with no previous history, then none of them can gain by deviating from DOC. In addition, in repeated games it is also important to ensure that, if at some point the game has a given history, a selfish station cannot take advantage of this history to gain profit by playing a strategy different from DOC. The following theorem confirms that *All-DOC* is a Nash equilibrium of any subgame (where a subgame is defined as the game resulting from starting to play with a certain history). Therefore, a selfish station cannot benefit by deviating from DOC independently of the previous history of the game.

**Theorem 6.** *All-DOC is a subgame perfect Nash equilibrium of the game.*

*Proof:* Since the proof of Theorem 5 is independent of the past history and can therefore be applied to any subgame, *All-DOC* is a Nash equilibrium of any subgame. ■

### *B. Multiple selfish stations*

The above results show the effectiveness of DOC against a single selfish station. In the following, we tackle the case when there are multiple selfish stations.

The following theorem shows that, by following a strategy different from DOC, multiple stations cannot gain any *aggregated* channel time.

**Theorem 7.** *Let us consider a scenario with $m$ selfish stations. If all other stations play DOC, the selfish stations cannot gain any aggregated channel time.*

*Proof:* Without loss of generality, let us consider that stations $i = \{1, \ldots, m\}$ are selfish. Applying a reasoning similar to Theorem 5 leads to

$$\sum_{i=1}^{m} \sum_{\Theta} t_i(\Theta) \leq m \sum_{\Theta} t^* \tag{86}$$

As the left hand side of the above equation is the aggregated channel time obtained by the selfish stations, and the right hand side is the aggregated channel time that they would obtain if the played DOC, this proves the theorem. ■

The above theorem shows that, if there is some selfish station that experiences a gain, this is because there is some other station that suffers a loss.

**Corollary 3.** *Let us consider a scenario with $m$ selfish stations. If all other stations play DOC and a selfish station $k$ receives a throughput larger than $r_k^*$, this means that there exists another selfish station $l$ that receives a throughput smaller than $r_l^*$ (where $r_k^*$ and $r_l^*$ are the throughputs obtained by stations $k$ and $l$ if all stations played DOC).*

*Proof:* If there is some station $k \in \{1, \ldots, m\}$ for which $r_k > r_k^*$, then we have that this station receives more channel time than it would receive if all stations played DOC. Since, according to Theorem 7, the selfish stations cannot gain any aggregated channel time, this means that there must necessarily be some other station $l \in \{1, \ldots, m\}$ that receives less channel time. This implies that $r_l < r_l^*$, which proves the corollary. ■

Based on the above, we argue that DOC is effective against multiple selfish stations, since two or more selfish stations cannot *simultaneously* gain profit and therefore do not have an incentive to play a coordinated strategy different from DOC.

## VI. Performance Evaluation

In this section we evaluate DOC by means of simulation to show that $(i)$ in the absence of selfish stations, DOC provides optimal performance while behaving stably and reacting quickly to changes, and $(ii)$ selfish stations cannot benefit by following a strategy different from DOC.

Unless otherwise stated, we assume that different observations of the channel conditions are independent, and the available transmission rate for a given SNR is given by the Shannon channel capacity:

$$R(h) = W log_2(1 + \rho |h|^2) \quad \text{bits/s} \tag{87}$$

where $W$ is the channel bandwidth, $\rho$ is the normalized average SNR and $h$ is the random gain of Rayleigh fading.

We implemented the DOC algorithm in OMNET++[6]. In the simulations, we set $W = 10^7$, $\mathcal{T}/\tau = 10$ and the interval of the controller $T_{total} = 10^5\tau$. For all results, 95% confidence intervals are below 0.5%.

### *A. Throughput evaluation*

For the throughput evaluation, we compare the performance of DOC to the following approaches: $(i)$ the static optimal configuration obtained in Section II (‘*static configuration*’), $(ii)$ the configuration proposed in [3] (‘*DOS*’), and $(iii)$ an approach that does not perform opportunistic scheduling but always transmits after successful contention (‘*non-opportunistic*’).

We consider a scenario with $N = 10$ stations, half of them with a normalized SNR of $\rho_1 = 1$ and the other half with a normalized SNR $\rho_2$ that varies from 1 to 10. Fig. 5 shows $\sum_i log(r_i)$, the metric that proportional fairness aims at maximizing, as a function of $\rho_2$. We observe that DOC performs at the same level as the benchmark given by the *static configuration*, while the other two approaches (*DOS* and *non-opportunistic*) provide a substantially lower performance.

For above scenario with $\rho_2 = 4$, Fig. 6 depicts the individual throughput allocation of two stations (where $r_1$ is the throughput of a station with $\rho_1$ and $r_2$ that of a station with $\rho_2$). DOC is effective in driving the system to the optimal point of operation and provides the same throughput as the *static configuration*. In contrast, the *DOS* approach exhibits a high degree of unfairness and provides the station with high SNR with a much higher throughput. The *non-opportunistic* approach provides a good level of fairness but has lower throughput due to the lack of opportunistic scheduling. In conclusion, the proposed DOC algorithm provides a good tradeoff between overall throughput and fairness.

[6] http://www.omnetpp.org/

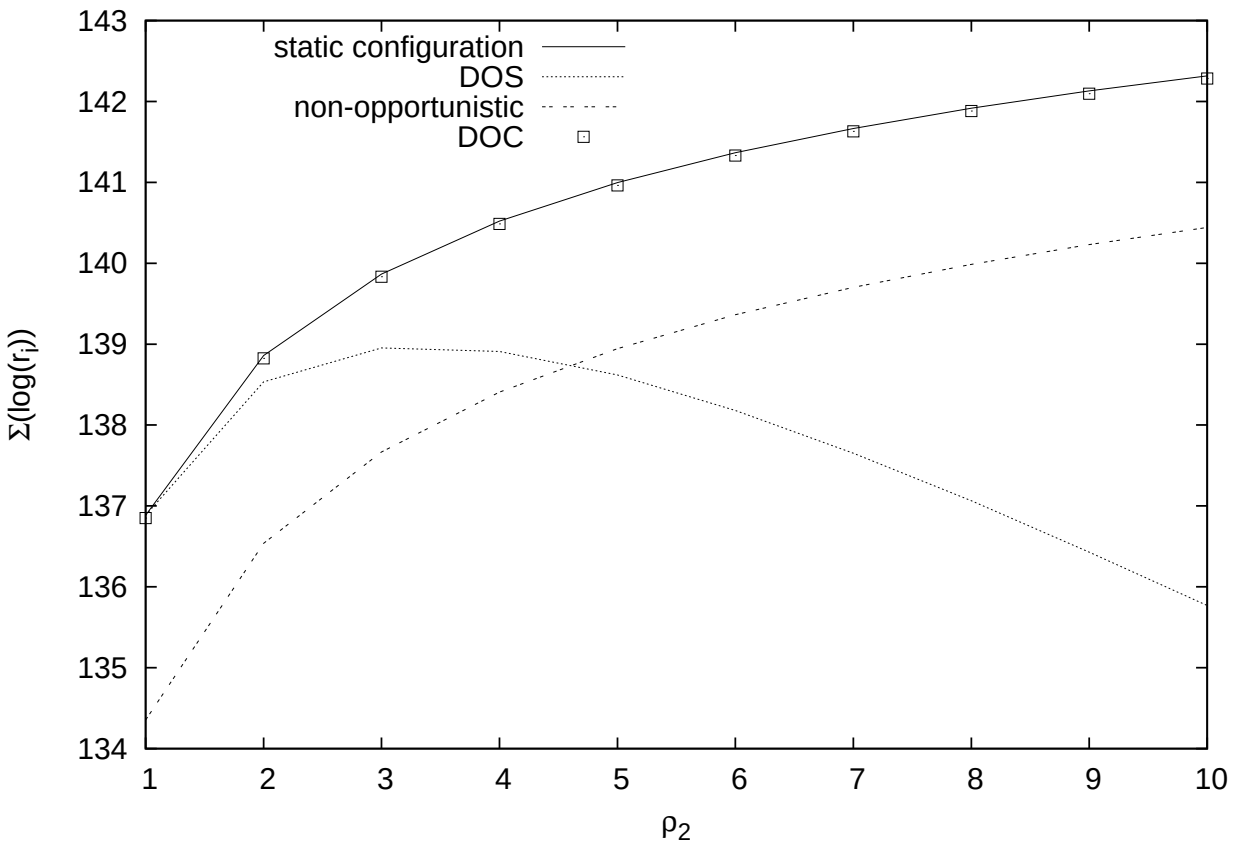

Fig. 5. Proportional fairness as a function of SNR ($\rho_1 = 1, 1 \leq \rho_2 \leq 10$).

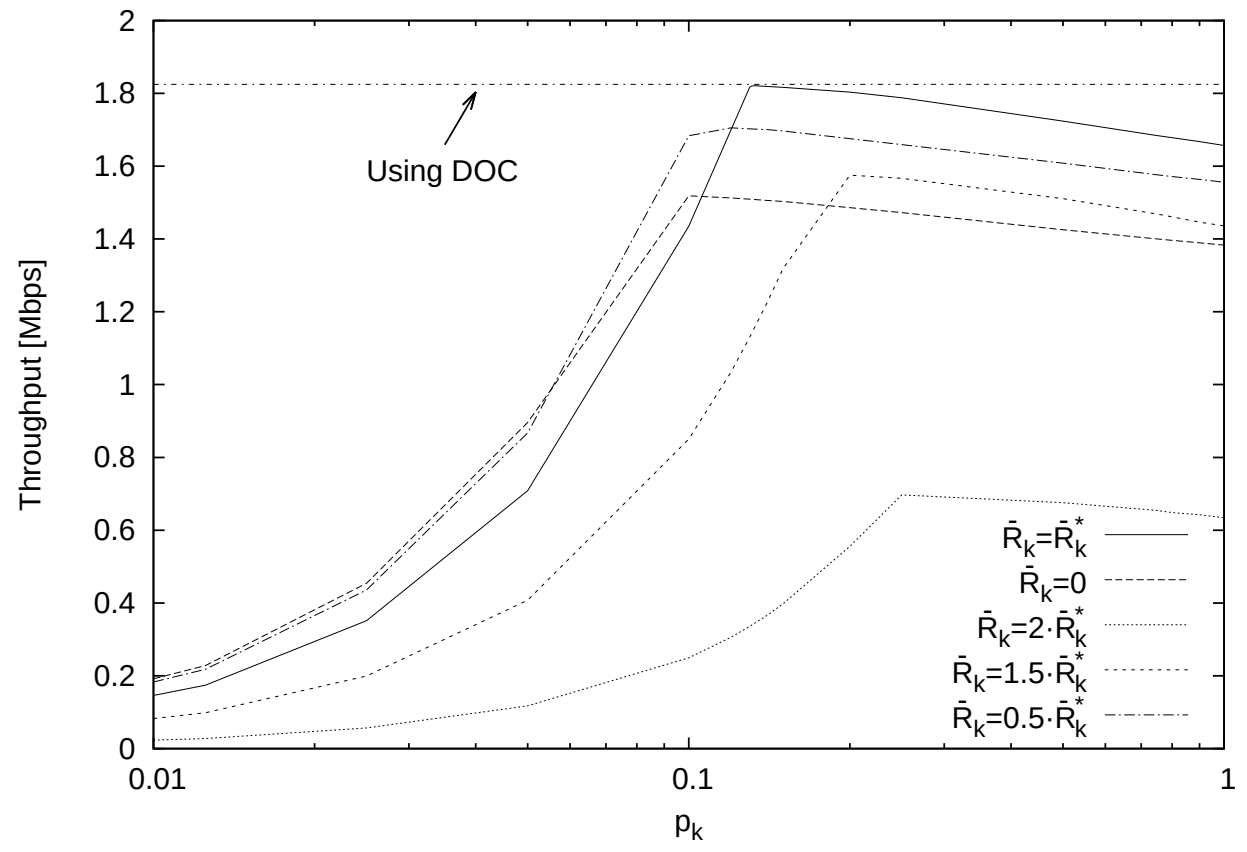

Fig. 7. Throughput of a selfish station for fixed configurations of $\{p_k, \bar{R}_k\}$.

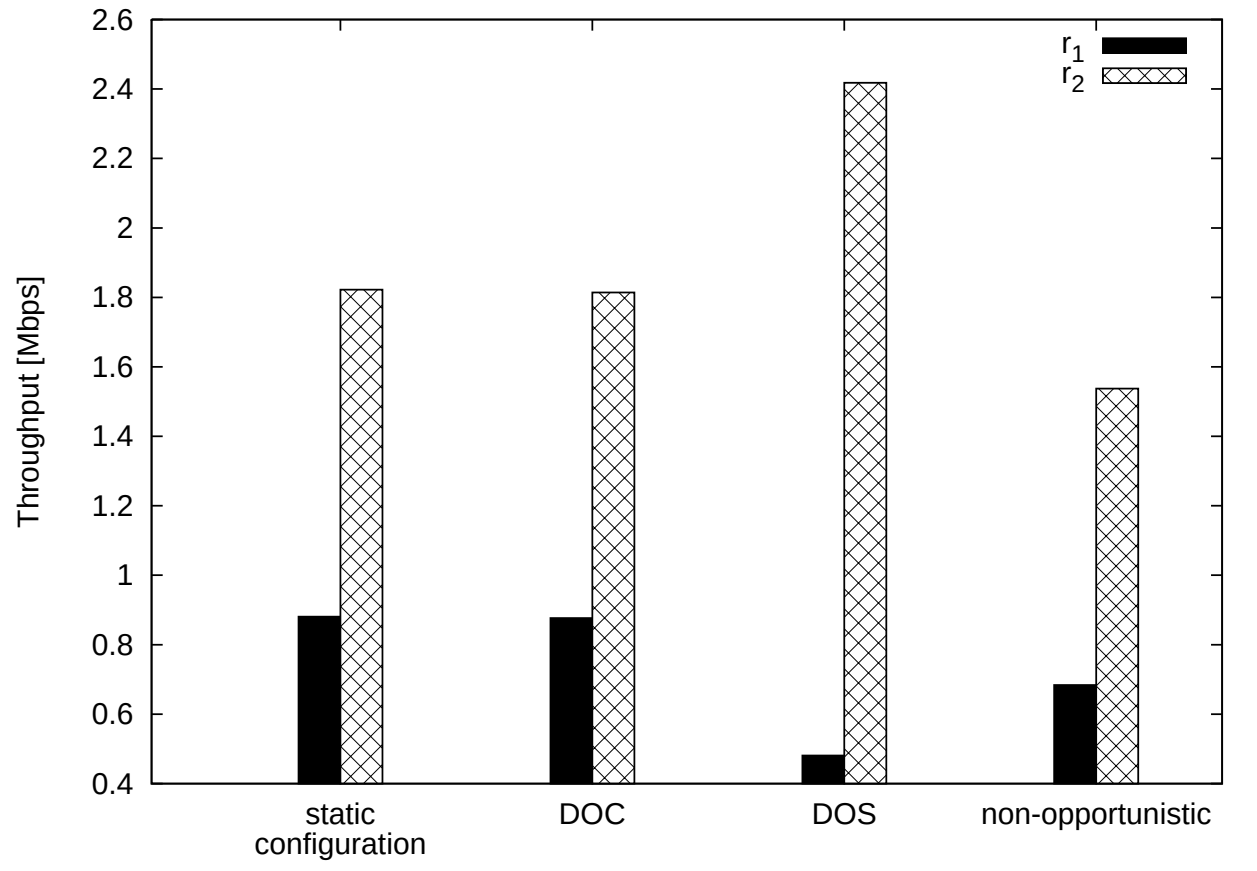

Fig. 6. Throughput for heterogeneous SNRs ($\rho_1 = 1, \rho_2 = 4$).

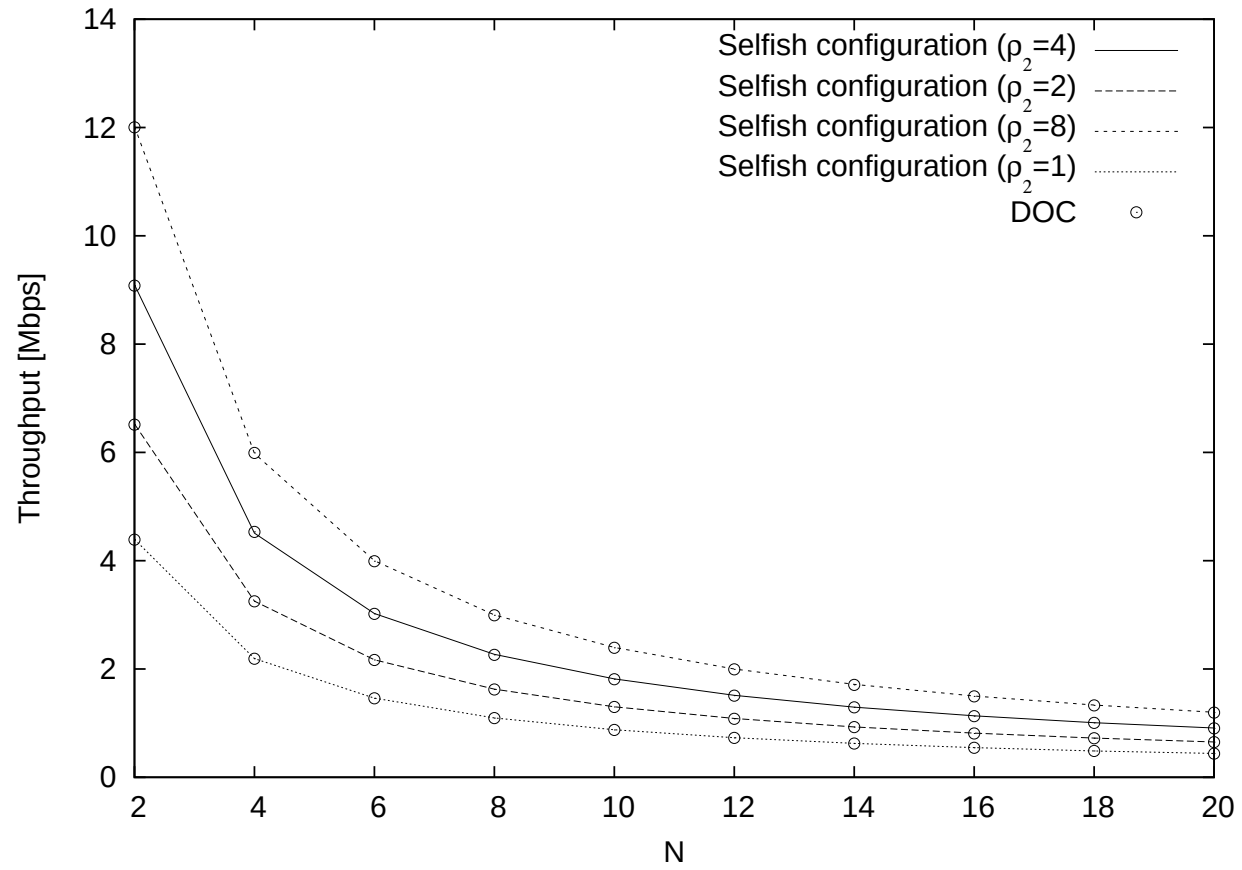

Fig. 8. Selfish station with fixed configuration for different $N$ and $\rho_2$ values.

### B. Selfish station with fixed configuration

We verify that a station cannot obtain more throughput with a selfish configuration than by playing DOC in a scenario with $N = 10$ stations, 5 of them with $\rho_1 = 1$ and the other half (including the selfish station) with $\rho_2 = 4$. The selfish station uses a fixed configuration and all other stations implement DOC. Fig. 7 shows the throughput of the selfish station for different $\{p_k, \bar{R}_k\}$ configurations of the selfish station. This is compared to the throughput that the station would obtain if it played DOC, given by the horizontal line.

We observe that none of the selfish configurations provides more throughput than DOC. Furthermore, $r_k$ is far from $r_k^*$ for $p_k < p_k^*$ and close to $r_k^*$ for $p_k > p_k^*$. This is a consequence of the design of $F_i$ in Section III.E. For $p_k < p_k^*$, the access probabilities of the honest stations satisfy $p_i < p_i^*$. With these values of $\mathbf{p}$, $F_i$ takes negative values that are large in absolute terms, which means that, according to Eq. (27), the selfish station receives much less channel time than the other stations and hence a throughput far from $r_k^*$. For $p_k > p_k^*$, we have $p_i > p_i^*$. These $\mathbf{p}$ lead to $F_i$ values that are close to the upper bound and, as the upper bound corresponds to $t_k = t^*$, this gives a throughput close to $r_k^*$ for the selfish station. In Section VI-E we show that this design leads to a robust behavior against selfish stations and transient conditions.

Fig. 8 analyzes the impact of fixed selfish configurations for a range of different $N$ and $\rho_2$ values. It shows the largest throughput that a selfish station can receive with a fixed configuration, which is obtained by performing an exhaustive search over the $\{p_k, \bar{R}_k\}$ space. This throughput is compared against the one that the station would receive if it played DOC. Again we observe that the station never benefits from playing selfishly, which validates the design of the DOC algorithm.

### C. Selfish station with variable configuration

According to Theorem 5, a selfish station cannot benefit from changing its configuration over time. For verification, we evaluate the throughput obtained by a selfish station with different adaptive strategies. These strategies are inspired by the schemes used in [20], [22] for a similar purpose. The underlying principle of all of them is that the cheating station uses a selfish configuration to gain throughput and, when it realizes that it is not gaining throughput, it assumes that it has been detected as selfish and switches back to the honest configuration to avoid being punished.

In particular, we consider the following strategies. The '*adaptive* $p_k$ *strategy*' fixes the $\bar{R}_k$ configuration of the selfish station to its optimal value, $\bar{R}_k = \bar{R}_k^*$, and modifies the $p_k$ configuration as follows: the station uses a selfish configuration of $p_k = 1$ as long as it obtains some gain, i.e. $r_k > r_k^*$. When $r_k$ drops below $r_k^*$, the station switches to the honest configuration, $p_k = p_k^*$, and stays with this configuration as long as $r_k$ stays below $0.95r_k^*$. It switches back to $p_k = 1$ when $r_k$ grows above $0.95r_k^*$. The '*adaptive $\bar{R}_k$ strategy*' fixes the $p_k$ configuration to the optimal value, $p_k = p_k^*$, and modifies the $\bar{R}_k$ configuration following a strategy similar to the one above: the station uses a selfish configuration of $\bar{R}_k = 0$ (i.e., it uses all transmission opportunities) as long as it obtains some gain and switches to the honest configuration when it stops benefiting. Finally, the '*adaptive $p_k$ and $\bar{R}_k$ strategy*' follows a similar behavior to the previous ones but adapts both the $p_k$ and the $\bar{R}_k$ configuration.

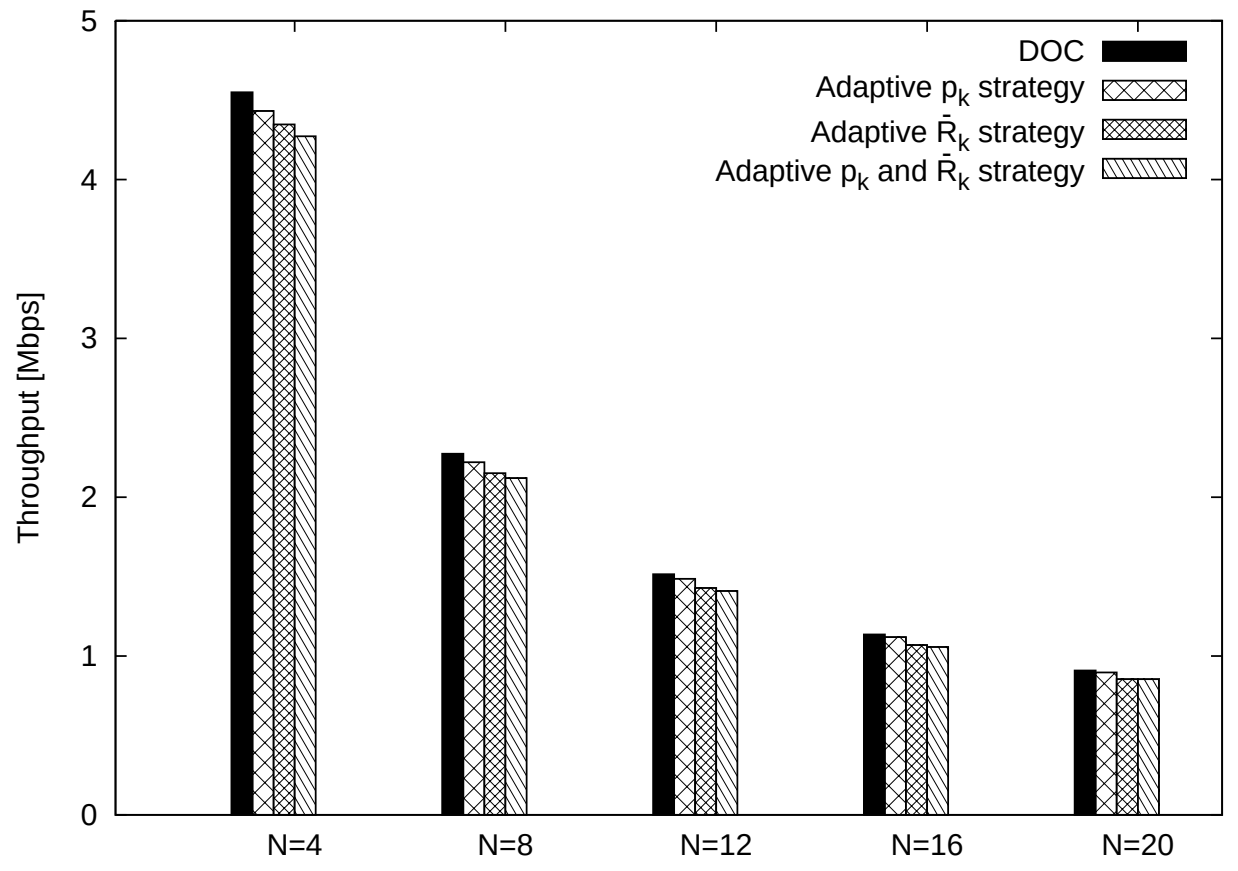

Fig. 9. Throughput of selfish station with different adaptive strategies.

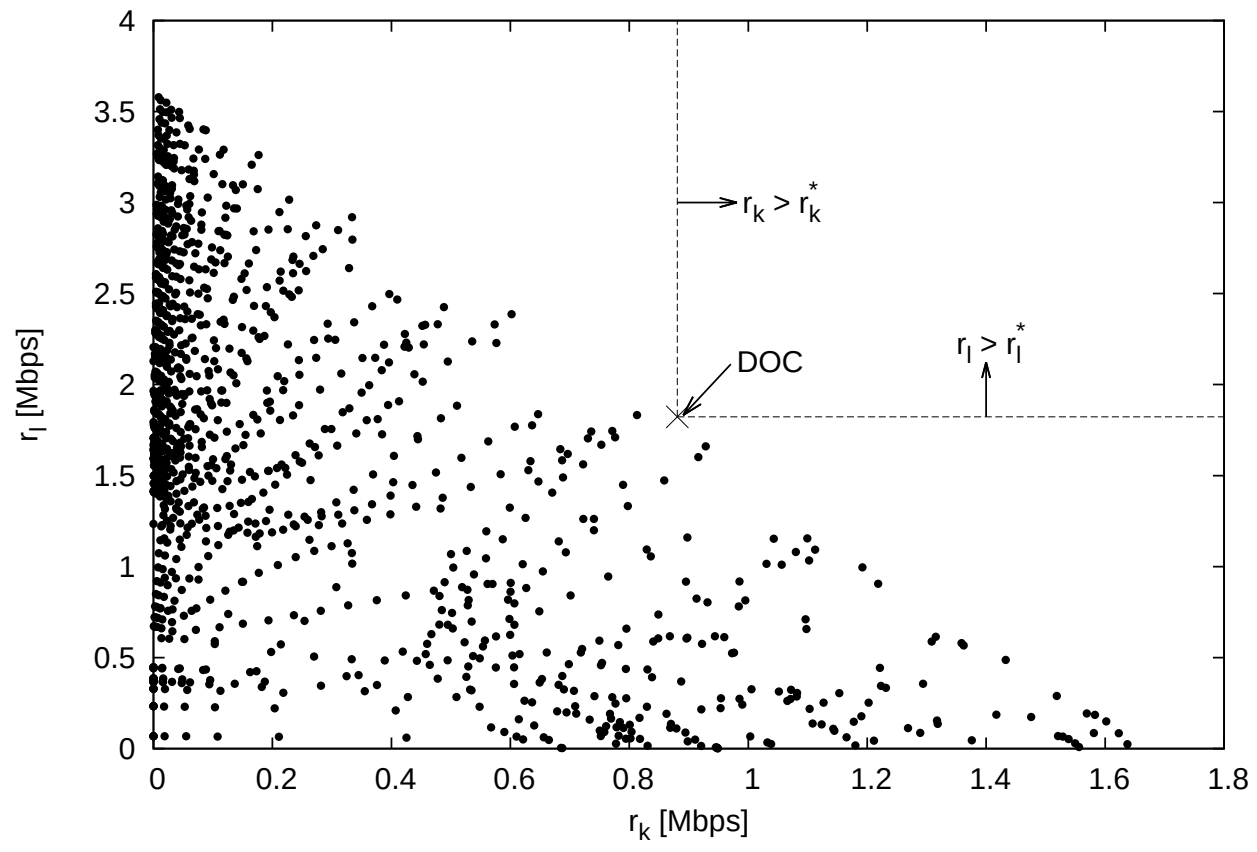

Fig. 10. Throughput obtained by multiple selfish stations.

Fig. 9 compares the throughput obtained with each of the above strategies against the one with DOC for different values of $N$. As expected, when all other stations play DOC, a given station maximizes its payoff playing DOC as well, as it obtains a larger throughput than with any of the other strategies, confirming the result of Theorem 5.

### D. Multiple selfish stations

Corollary 3 states that multiple selfish stations cannot simultaneously benefit by deviating from DOC, as it is only possible that one or more of the selfish stations experience some throughput gains if there are some other selfish stations that suffer some loss.

To validate the result, we consider a network with $N = 10$ stations including two selfish stations, half of them (including one of the selfish stations) with $\rho_1 = 1$ and the other half (including the other selfish station) with $\rho_2 = 4$. We perform an exhaustive search over a wide range of $\{p_i, \bar{R}_i\}$ configurations of the two selfish stations. The results of this experiment are depicted in Fig. 10, which shows the throughput obtained by the two selfish stations ($r_k$ and $r_l$) for each of the configurations used in the exhaustive search. The figure also shows the throughput of the two stations when they both play DOC. There is no configuration that simultaneously improves the throughput of the two selfish stations, which confirms the result of Corollary 3.

We also observe from the figure that the region of feasible allocations has a tringular shape. This is a consequence of Theorem 7: since the maximum aggregated channel time that the two stations can obtain is fixed, any throughput increase in one station leads to a decrease in the other station of the same amount scaled by a constant factor that depends on the respective radio conditions.

### E. Robustness to selfish behavior and transient conditions

For a setting similar to that of Fig. 7 with 10 stations, half of them with $\rho_1 = 1$ and half with $\rho_2 = 4$, and one selfish station with a fixed configuration and $\rho_2 = 4$, we investigate the overall throughput of the wireless system. Again, the throughput obtained when all stations play DOC is given by the horizontal line. From Fig. 11 we see that the overall throughput is close to optimal for low values of the access probability $p_k$ of the selfish station and only gradually decreases for high values of $p_k$. For low values of $p_k$, well-behaved stations contend with a higher access probability than the selfish station which yields an almost optimal throughput. For high values of $p_k$, the selfish station has to be punished which unavoidably results in some throughput loss. However, the level of punishment is minimized to avoid driving the collision probability to unnecessarily high levels that harm the overall throughput. Hence, even for very high $p_k$ and the subsequent high rate of contention collisions, some throughput remains for the well-behaved stations (note from Fig. 7 that the maximum throughput of the selfish station is less than $\sim$1.8 Mbps). We conclude that, as intended, the design of $F_i$ maintains a level of throughput as high as possible for the well-behaved stations.

In addition to robustness against selfish behavior (as seen above), our design of $F_i$ also aims at providing robustness in transient conditions. We investigate this through the following experiment: in a wireless network with 10 stations, a new station joins every 100 intervals, with its $p_i$ initially set to 0.5, stays for 50 intervals and then leaves the system. With our design of $F_i$ the total throughput obtained in this scenario is equal to 9.67 Mbps, while with a design of $F_i$ 10 times smaller, it is only 6.52 Mbps, confirming the robustness of DOC to transient network conditions.

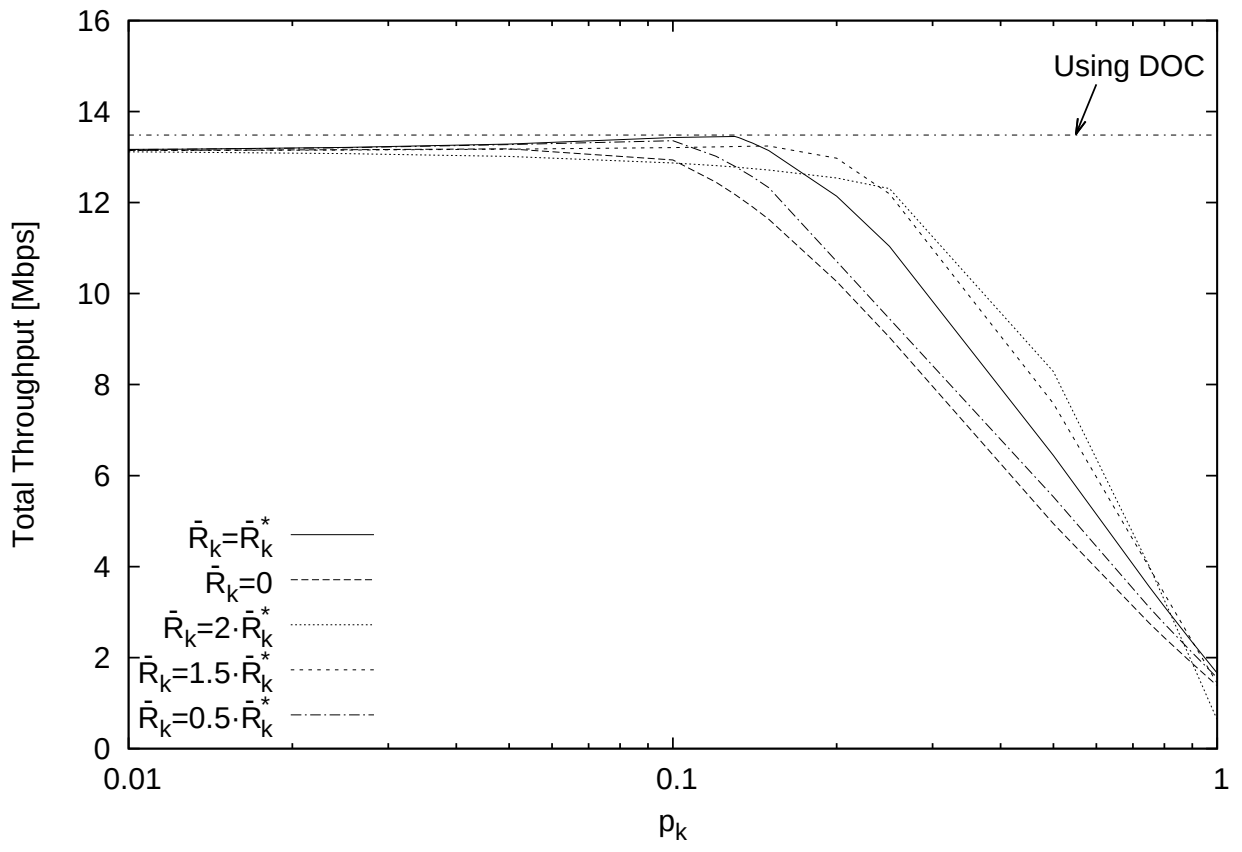

Fig. 11. Total system throughput in the presence of a selfish station.

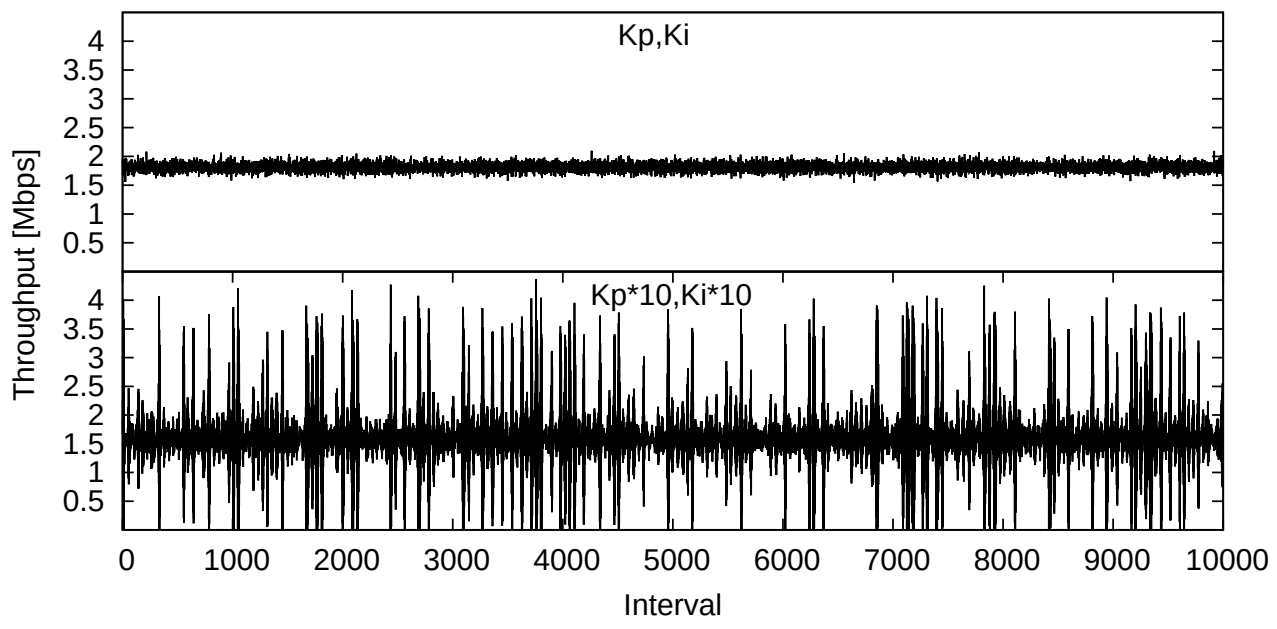

Fig. 12. Stability analysis of the parameters of the PI controller.

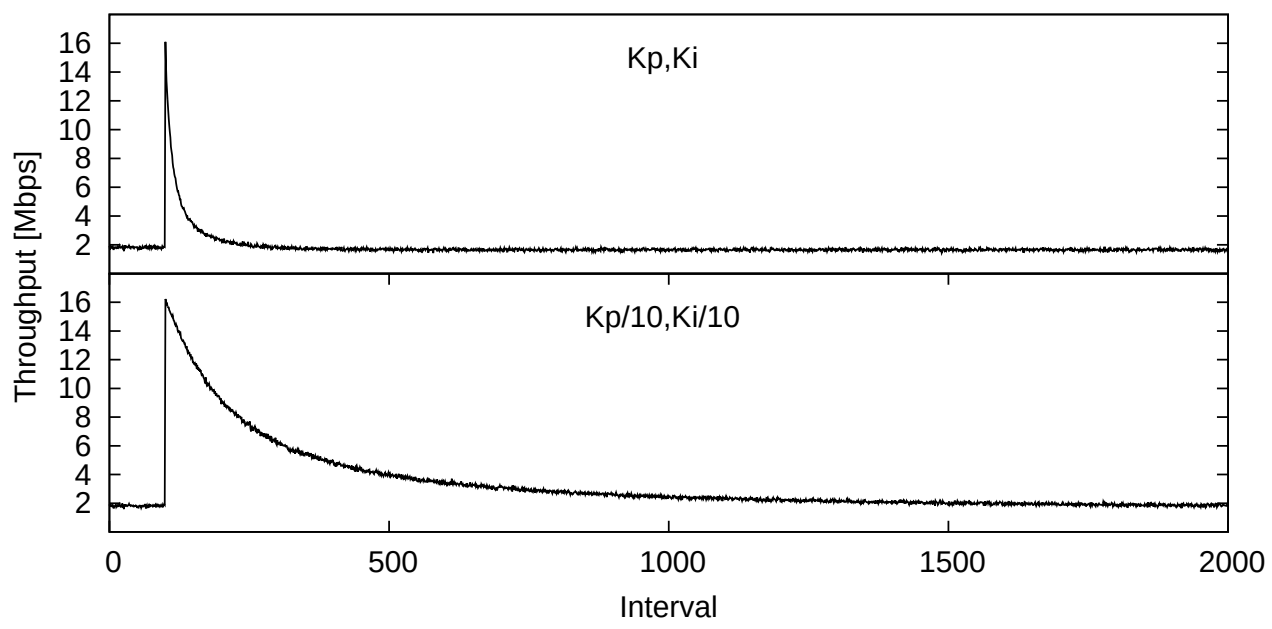

Fig. 13. Speed of reaction provided by the parameters of the PI controller.

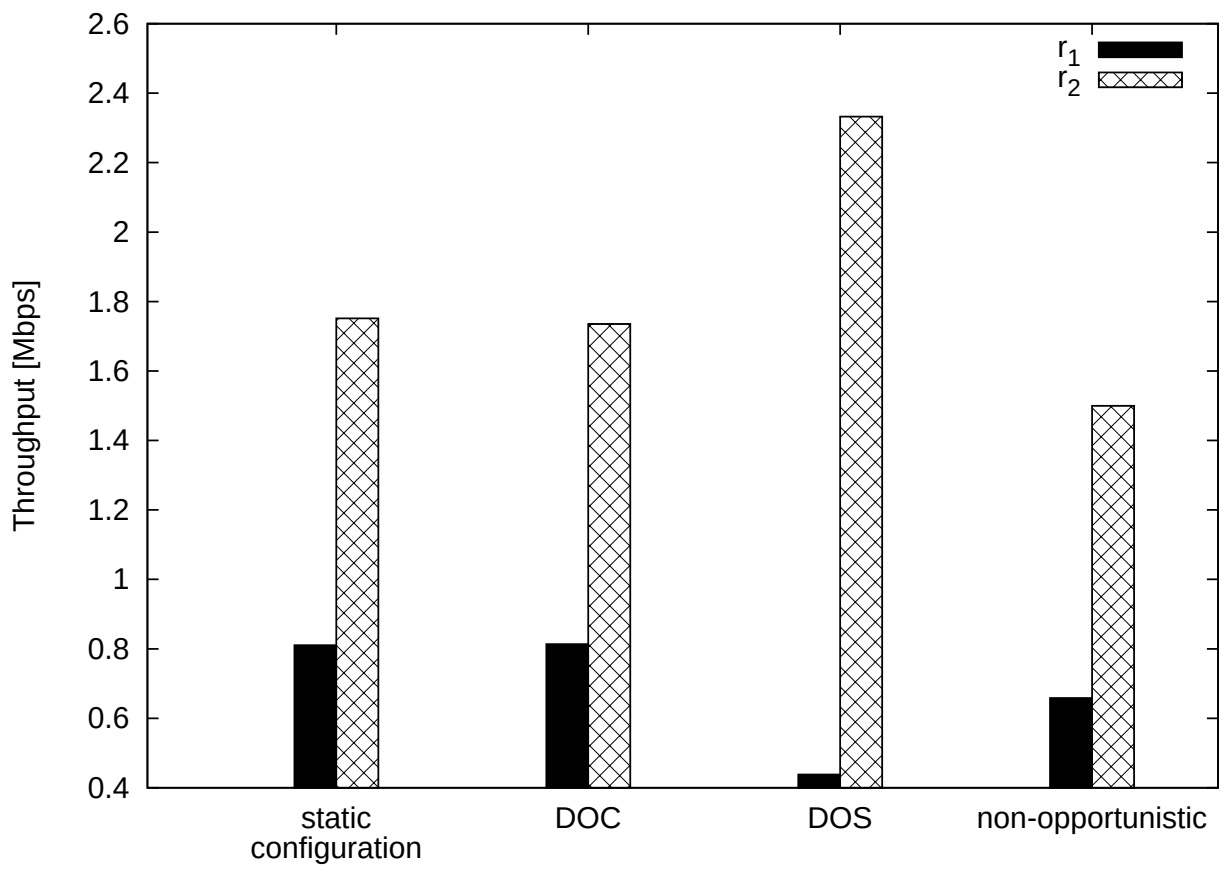

Fig. 14. Performance with *Jakes' channel model*.

### F. Parameter setting of the PI controller

The main objective in the setting of the $K_p$ and $K_i$ parameters proposed in Section IV is to achieve a good tradeoff between stability and speed of reaction.

To validate that our system guarantees a stable behavior, we analyze the evolution over time of the throughput received by a station for the chosen $\{K_p, K_i\}$ setting and a configuration of these parameters 10 times larger, in a wireless network with $N = 10$ stations. We observe from Fig. 12 that with the proposed setting (labeled "$K_p, K_i$"), the throughput shows only minor deviations around its average value, while for a larger setting (labeled "$K_p * 10, K_i * 10$"), it shows unstable behavior with drastic oscillations.

To investigate the speed with which the system reacts against selfish stations, we use a wireless network with $N = 10$ stations where initially all stations play DOC and, after 50 intervals, one station turns selfish and changes its access probability to $p_k = 1$. Fig. 13 shows the evolution of the throughput of the selfish station over time. We observe from the figure that with our setting (labeled "$K_p, K_i$"), the system reacts quickly, and after a few tens of intervals the selfish station no longer benefits from its behavior. In contrast, for a setting of these parameters 10 times smaller (labeled "$K_p/10, K_i/10$"), the reaction is very slow and it takes almost 2000 intervals until the station stops benefiting from its misbehavior.

The results show that with a larger setting of $\{K_p, K_i\}$ the system suffers from instability while with a smaller one it reacts too slowly. Hence, the proposed setting provides a good tradeoff between stability and speed of reaction.

### G. Impact of channel coherence time

Our channel model is based on the assumption that different observations of the channel conditions are independent. In order to understand the impact of this assumption, we repeat the experiment of Fig. 6 using *Jakes' channel model* [23] to obtain the different channel observations. The results, for a Doppler frequency of $f_D = 2\pi/100\tau$, are given in Fig. 14. We observe that the throughput obtained is slightly smaller than that of Fig. 6. This is due to the fact that when the channel is bad, a station does not transmit after a successful contention and therefore it takes (on average) a shorter time until the next successful contention of this station. As a result, a station accesses more often the channel when it is bad than when it is good, which introduces a bias that slightly reduces the throughput. Overall, the results are sufficiently similar to those of Fig. 6 to conclude that our assumption on the channel model only has a minor impact on the resulting performance.

We further investigate whether, in the above scenario, a station with $\rho_2 = 4$ could obtain more throughput by using a selfish configuration. While the station obtains 1.752 Mbps with DOC, it can obtain up to 1.757 Mbps with a selfish configuration. Note that this increase is not due to the DOC design, as no configuration gives more channel time to the selfish station, but rather due to the fact that the transmission rate threshold of [3] is not truly optimal under *Jakes' channel model*. In any case, the throughput gain of the selfish station is negligible.

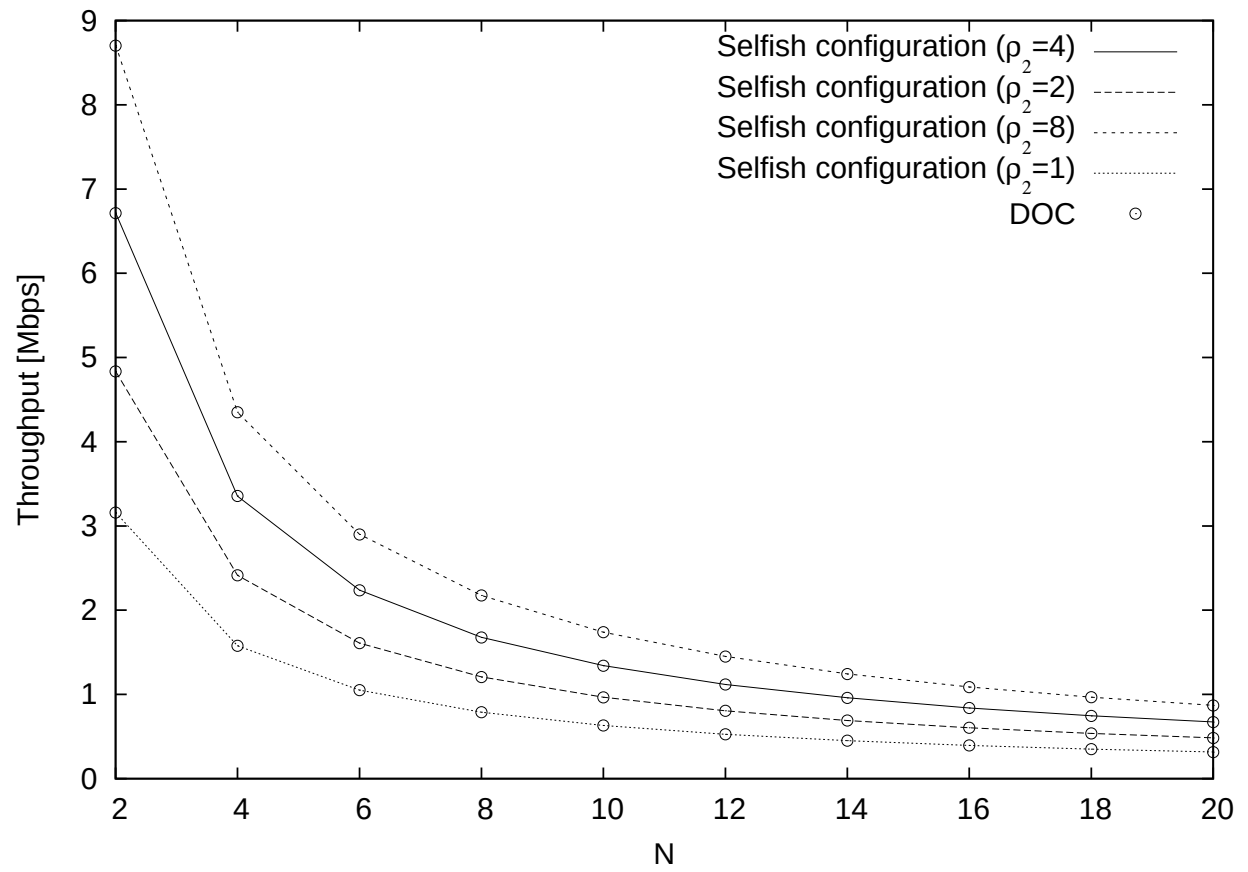

Fig. 15. Throughput comparison for a discrete set of rates.

### *H. Discrete set of transmission rates*

While all previous experiments assumed continuous rates, our analysis as well as the design of the DOC algorithm does not rely on any assumption on the mapping of SNR to transmission rates and therefore works for any (continuous or discrete) mapping function. To show that DOC is effective when only a set of discrete rates is allowed, we analyze a wireless system in which the only transmission rates available are $\{1, 2, 5.5, 12, 24, 48, 54\}$ Mbps. For a given SNR, we choose the largest available transmission rate that is smaller than the one given by Eq. (87).

We repeat the experiment of Fig. 9 with discrete rates, and compare the throughput of a selfish station against the throughput that this station obtains when it plays DOC. The results in Fig. 15 confirm that a station cannot benefit from playing selfish. We further observe that, as expected, throughputs are smaller than those of Fig. 9 since, with the discrete mapping of SNR to rates, smaller transmission rates are achieved on average.

## VII. Conclusions

Recently proposed Distributed Opportunistic Scheduling (DOS) techniques provide throughput gains in wireless networks that do not have a centralized scheduler. One of the problems of these techniques is, however, that they are vulnerable to malicious users which may configure their parameters to obtain a greater share of the wireless resources at the expense of other, well-behaved, users. In this paper we address the problem by proposing a novel algorithm that prevents such throughput gains from selfish behavior.

With our approach, upon detecting a selfish user, stations react by using a more aggressive parameter configuration which serves to punish the selfish station. Such an adaptive algorithm has to carefully adjust the reaction against a selfish station to avoid that the system turns unstable by overreacting. A key aspect of the paper is that we use of tools from the fields of *multivariable control theory* combined with game theory in the design of our algorithm.

We conducted a control theoretic analysis of the DOC algorithm that shows that, when all the stations in the wireless network run DOC, the system behaves stably and converges to the desired configuration. We then used this control theoretic analysis to find a setting that provides a good tradeoff between stability and speed of reaction. In addition, we performed a game theoretic analysis of DOC based on *repeated games* to evaluate its behavior when there are one or more selfish stations in the wireless network. The analysis shows that neither a single selfish station nor several cooperating selfish stations can benefit from playing a strategy different from DOC, and that this holds for fixed as well as for adaptive strategies. Furthermore, the DOC strategy represents a subgame perfect Nash equilibrium.

## References

[1] M. Andrews *et al.*, “Providing quality of service over a shared wireless link,” *IEEE Communications Magazine*, vol. 39, no. 2, February 2001.

[2] P. Viswanath, D. N. Tse, and R. Laroia, “Opportunistic beamforming using dumb antennas,” *IEEE Transactions on Information Theory*, vol. 48, no. 6, pp. 1277–1294, June 2002.

[3] D. Zheng, W. Ge, and J. Zhang, “Distributed opportunistic scheduling for ad hoc networks with random access: an optimal stopping approach,” *IEEE Transactions on Information Theory*, vol. 55, no. 1, January 2009.

[4] D. Zheng, , M.-O. P. W. Ge, H. V. Poor, and J. Zhang, “Distributed opportunistic scheduling for ad hoc communications with imperfect channel information,” *IEEE Transactions on Wireless Communications*, vol. 7, no. 12, pp. 5450 – 5460, December 2008.

[5] P. Thejaswi, J. Zhang, M.-O. Pun, H. V. Poor, and D. Zheng, “Distributed opportunistic scheduling with two-level probing,” *IEEE/ACM Transactions on Networking*, vol. 18, no. 5, October 2010.

[6] S. Tan, D. Z. J. Zhang, and J. R. Zeidler, “Distributed opportunistic scheduling for ad-hoc communications under delay constraints,” in *Proceedings of IEEE INFOCOM*, San Diego, CA, March 2010.

[7] F. Kelly, “Charging and rate control for elastic traffic,” *European Transactions on Telecommunications*, vol. 8, pp. 33–37, 1997.

[8] G. Holland, N. Vaidya, and P. Bahl, “A rate-adaptive mac protocol for multi-hop wireless networks,” in *Proceedings of ACM MOBICOM*, Rome, Italy, July 2001.

[9] B. Sadhegi, V. Kanodia, A. Sabharwal, and E. Knightly, “Opportunistic media access for multirate ad hoc networks,” in *Proceedings of ACM MOBICOM*, Atlanta, GA, September 2002.

[10] P. Gupta, Y. Sankarasubramaniam, and A. Stolyar, “Random-access scheduling with service differentiation in wireless networks,” in *Proceedings of IEEE INFOCOM*, Miami, FL, March 2005.

[11] T. Glad and L. Ljung, *Control theory: multivariable and nonlinear methods*. Taylor & Francis, 2000.

[12] P. Patras, A. Banchs, P. Serrano, and A. Azcorra, “A Control Theoretic Approach to Distributed Optimal Configuration of 802.11 WLANs,” *IEEE Transactions on Mobile Computing*, vol. 10, no. 6, June 2011.

[13] C. Hollot, V. Misra, D. Towsley, and W. B. Gong, “On designing improved controllers for AQM routers supporting TCP flows,” in *Proceedings of IEEE INFOCOM*, Anchorage, Alaska, April 2001.

[14] G. Boggia, P. Camarda, L. A. Grieco, and S. Mascolo, “Feedback-based control for providing real-time services with the 802.11e mac,” *IEEE/ACM Transactions on Networking*, vol. 15, no. 2, April 2007.

[15] K. Aström and R. M. Murray, *Feedback Systems*. Princeton University Press, 2008.

[16] C. V. Hollot, V. Misra, D. Towsley, and W. B. Gong, “A Control Theoretic Analysis of RED,” in *Proceedings of IEEE INFOCOM*, Anchorage, Alaska, April 2001.

[17] K. Aström and B. Wittenmark, *Computer-controlled systems, theory and design*, 2nd ed. Prentice Hall International Editions, 1990.

[18] G. F. Franklin, J. D. Powell, and M. L. Workman, *Digital Control of Dynamic Systems*, 2nd ed. Addison-Wesley, 1990.

[19] D. Fudenberg and J. Tirole, *Game Theory*. MIT Press, 1991.

[20] M. Cagalj, S. Ganeriwal, I. Aad, and J.-P. Hubaux, “On selfish behavior in csma/ca networks,” in *Proceedings of IEEE INFOCOM*, Miami, Florida, March 2005.

[21] J. Konorski, “A game-theoretic study of csma/ca under a backoff attack,” *IEEE/ACM Transactions on Networking*, vol. 16, no. 6, December 2006.

[22] L. Buttyan and J.-P. Hubaux, *Security and Cooperation in Wireless Networks*. Cambridge: Cambridge University Press, 2008.

[23] W. C. Jakes, *Microwave Mobile Communications*. New York: John Wiley & Sons Inc., 1975.